\documentclass[notitlepage,twocolumn,showpacs,preprintnumbers,amsmath,amssymb,superscriptaddress, pra,floatfix]{revtex4-1}
\usepackage{bm}
\usepackage{graphicx}
\usepackage{dcolumn}
\usepackage{tabularx}
\usepackage{bm}
\usepackage{epstopdf}
\usepackage{amsmath}
\usepackage{color}
\usepackage{here}
\allowdisplaybreaks

\begin{document}
\preprint{Preprint}
\title{Photon correlation measurements of stochastic limit cycles emerging from high-$Q$ nonlinear silicon photonic crystal microcavities}
\author{N. Takemura}
\email[E-mail: ]{naotomo.takemura.ws@hco.ntt.co.jp}
\affiliation{NTT Nanophotonics Center, NTT Corp., 3-1, Morinosato Wakamiya Atsugi, Kanagawa 243-0198, Japan}
\affiliation{NTT Basic Research Laboratories, NTT Corp., 3-1, Morinosato Wakamiya Atsugi, Kanagawa 243-0198, Japan}
\author{M. Takiguchi}
\affiliation{NTT Nanophotonics Center, NTT Corp., 3-1, Morinosato Wakamiya Atsugi, Kanagawa 243-0198, Japan}
\affiliation{NTT Basic Research Laboratories, NTT Corp., 3-1, Morinosato Wakamiya Atsugi, Kanagawa 243-0198, Japan}
\author{H. Sumikura}
\affiliation{NTT Nanophotonics Center, NTT Corp., 3-1, Morinosato Wakamiya Atsugi, Kanagawa 243-0198, Japan}
\affiliation{NTT Basic Research Laboratories, NTT Corp., 3-1, Morinosato Wakamiya Atsugi, Kanagawa 243-0198, Japan}
\author{E. Kuramochi}
\affiliation{NTT Nanophotonics Center, NTT Corp., 3-1, Morinosato Wakamiya Atsugi, Kanagawa 243-0198, Japan}
\affiliation{NTT Basic Research Laboratories, NTT Corp., 3-1, Morinosato Wakamiya Atsugi, Kanagawa 243-0198, Japan}
\author{A. Shinya}
\affiliation{NTT Nanophotonics Center, NTT Corp., 3-1, Morinosato Wakamiya Atsugi, Kanagawa 243-0198, Japan}
\affiliation{NTT Basic Research Laboratories, NTT Corp., 3-1, Morinosato Wakamiya Atsugi, Kanagawa 243-0198, Japan}
\author{M. Notomi}
\affiliation{NTT Nanophotonics Center, NTT Corp., 3-1, Morinosato Wakamiya Atsugi, Kanagawa 243-0198, Japan}
\affiliation{NTT Basic Research Laboratories, NTT Corp., 3-1, Morinosato Wakamiya Atsugi, Kanagawa 243-0198, Japan}
\date{\today}
\begin{abstract}
We performed measurements of photon correlation [$g^{(2)}(\tau)$] in driven nonlinear high-$Q$ silicon (Si) photonic crystal (PhC) microcavities. The measured $g^{(2)}(\tau)$ exhibits damped oscillatory behavior when input pump power exceeds a critical value. From comparison between experiments and simulations, we attribute the measured oscillation of $g^{(2)}(\tau)$ to self-pulsing (a limit cycle) emerging from an interplay between photon, carrier, and thermal dynamics. Namely, the oscillation frequency of $g^{(2)}(\tau)$ corresponds to the oscillation period of the limit cycle, while its finite coherence (damping) time originates from the stochastic nature of the limit cycle. From the standpoint of phase reduction theory, we interpret the measured coherence time of $g^{(2)}(\tau)$ as the coherence (diffusion) time of a generalized phase of the limit cycle. Furthermore, we show that an increase in laser input power enhances the coherence time of $g^{(2)}(\tau)$ up to the order of microseconds, which could be a demonstration of the stabilization of a stochastic limit cycle through pumping. 
\end{abstract}
\maketitle
 
\begin{figure}
\includegraphics[width=0.48\textwidth]{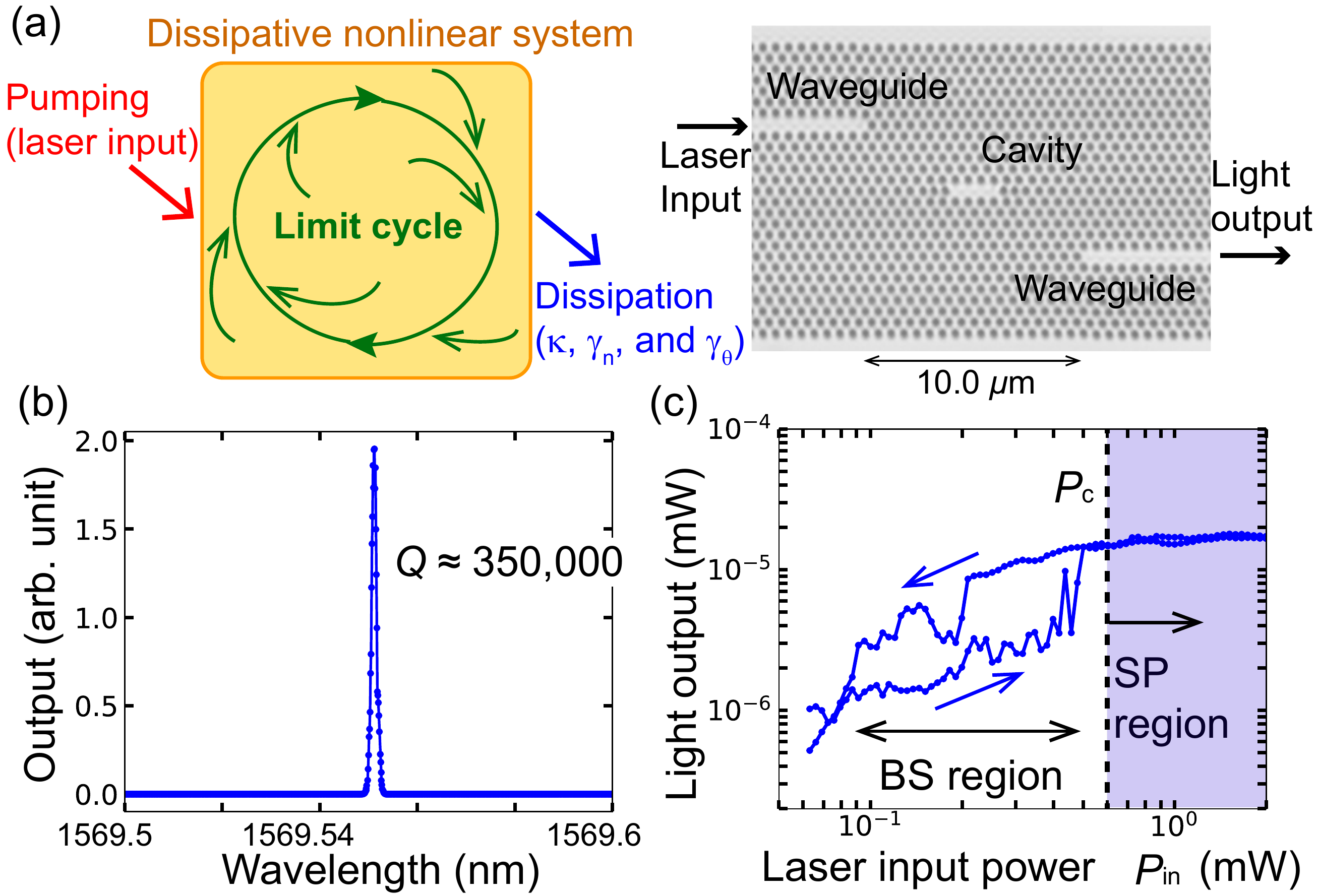}
\caption{(a) Illustration of a limit cycle emerging from a driven dissipative nonlinear system (left) and a laser scanning microscope image of the high-Q Si PhC microcavity (right). (b)  Laser transmission spectrum of the cavity showing the resonance of a fundamental mode. (c) Output intensity $I_{\rm out}$ as a function of input power $P_{\rm in}$, where $P_c$ is the critical laser input power for self-pulsing. BS and SP represent bistable and self-pulsing regions, respectively. In this paper, $P_{\rm in}$ is the fiber output power of the tunable semiconductor laser.}
\label{fig:schematic}
\end{figure}

A limit cycle is a universal natural phenomenon observed in a variety of systems ranging from electrical circuits to biological systems. In particular, in living systems, limit cycles play fundamental roles as, for example, biochemical oscillators, including cell cycles and circadian clock s \cite{Novak2008}. Importantly, as schematically shown on the left in Fig. \ref{fig:schematic}(a), limit cycles exist only for nonlinear dissipative systems, and they are qualitatively different from periodic oscillations in conservative systems such as simple pendulums. For example, the orbit of a pendulum is determined by the initial condition and becomes unstable with perturbation, whereas a limit cycle has a stable orbit, which is an attractor independent of an initial condition but controlled by system parameters such as pump power. At the same time, dissipative systems are usually noisy environments. Therefore, biochemical oscillators work as stochastic limit cycles, and strategies to maintain the precision of stochastic biochemical oscillators have been actively investigated in theoretical biophysics and biochemistry \cite{Gaspard2002,Gonze2002,Qian2006,Cao2015,Barato2016,Fei2018,Nguyen2018}. In this direction, a novel strategy is to increase the amplitude of a limit cycle \cite{Gaspard2002,Gonze2002}, which can be achieved by pumping \cite{Qian2006} or by free-energy dissipation \cite{Cao2015,Fei2018}. Furthermore, in discussing the precision of a limit cycle, a theoretical idea called ``phase reduction" proposed by Winfree and Kuramoto \cite{Kuramoto2003,Nakao2017} plays a key role, which reduces high-dimensional limit cycle dynamics to one-dimensional ``phase" dynamics along a limit cycle's orbit.
 
In this paper, we report experimental investigations of stochastic limit cycles in the optical domain and demonstrate a strategy for stabilizing a stochastic limit cycle with pumping. Our system is based on a driven silicon (Si) photonic crystal (PhC) high-$Q$ microcavity. A photonic microcavity device confines photons inside a nanoscale mode volume, which strongly enhances thermo-optic (TO) and carrier-induced optical nonlinearities in a medium such as a III-V material and Si \cite{Barclay2005,Uesugi2006,Leuthold2010}. Using the enhanced optical nonlinearity, optical bistability has been demonstrated with microcavities \cite{Tanabe2005,Notomi2005,Tanabe2007,Weidner2007,Haret2009,Rossi2009}. Furthermore, it is known that nonlinear photonic microcavities exhibit Hopf bifurcation, self-pulsing (a limit cycle) \cite{Priem2005,Johnson2006,Pernice2010,Malaguti2011,Cazier2013,Yacomotti2013,Yu2017}, and excitability \cite{Yacomotti2006,Brunstein2012}. In particular, in Refs. \cite{Yacomotti2006,Brunstein2012}, not only excitability but also limit cycle oscillation has been demonstrated in driven PhC cavities. Advantages of using a photonic system include controllability of pump power and dissipation and ease of measurements, such as real-time and photon correlation measurements. Therefore, photonic limit cycles could serve as artificial laboratories for understanding stochastic dynamical systems including biochemical oscillators. Furthermore, since optical limit cycles in all-Si PhC cavities operate at room temperature, it will be easy to integrate optical clocks in future silicon photonics circuits. In our study, we measured second-order photon correlation functions [$g^{(2)}(\tau)$] for a light output of the driven cavity. When laser input power exceeded a critical value, $g^{(2)}(\tau)$ exhibited damped oscillation. Together with numerical simulations, we show that the origin of the oscillation of $g^{(2)}(\tau)$ is self-pulsing (a limit cycle). Next, we argue that the finite coherence time of $g^{(2)}(\tau)$ originates from the stochastic nature of the system. The coherence time of $g^{(2)}(\tau)$ is interpreted as the coherence (diffusion) time of the generalized phase of the limit cycle \cite{Kuramoto2003,Nakao2017}. By measuring the input power dependence of the coherence time of $g^{(2)}(\tau)$, we observed an enhancement of the phase coherence time up to the order of microseconds with an increase in input power. Finally, we discuss the observed enhancement of the phase coherence time as a general property of a limit cycle, namely as a demonstration of stabilization of a stochastic limit cycle through pumping \cite{Gaspard2002,Gonze2002,Qian2006,Nguyen2018}. 
 
On right in Fig. \ref{fig:schematic}(a) is an image of our device, which is based on a 2D Si PhC slab with a cavity and two waveguides. The lattice constant, air-hole radius, and thickness of the PhC slab are 412, 100, and 215 nm, respectively. All the experiments were performed at room temperature. The cavity resonance of the fundamental mode is $\lambda_c=1569.55$ nm, and the $Q$ value is around 3.5$\times$10$^5$ [see Fig. \ref{fig:schematic}(b)]. The corresponding cavity photon lifetime, including losses to the waveguides, is around $1/\kappa=300$ ps. This very high $Q$ value was achieved by using the ultrahigh-Q design proposed in Ref. \cite{Kuramochi2014}, which omits three air holes and employs careful modulation of surrounding air holes [for further details about the device, see Section IA in the Supplemental Material (SM)]. We drive the cavity through the input waveguide with a tunable semiconductor laser, while we measure light outputs through the output waveguide. We introduce a normalized frequency detuning $\delta$ between the cavity resonance and laser input, which is defined as $\delta=(\omega_L-\omega_c)/\kappa$ with the cavity resonance frequency $\omega_c$, the laser input frequency $\omega_L$, and a field decay rate $\kappa$. In the measurements, we fixed the detuning as $\delta\simeq-2$. To measure second-order photon correlation functions [$g^{(2)}(\tau)$], we employed superconducting nanowire single-photon detectors (SNSPDs) and a conventional start-stop Hanbury Brown-Twiss (HBT) interferometer. For real-time measurements, we used an avalanche photodiode (APD). 
\begin{figure}
\includegraphics[width=0.48\textwidth]{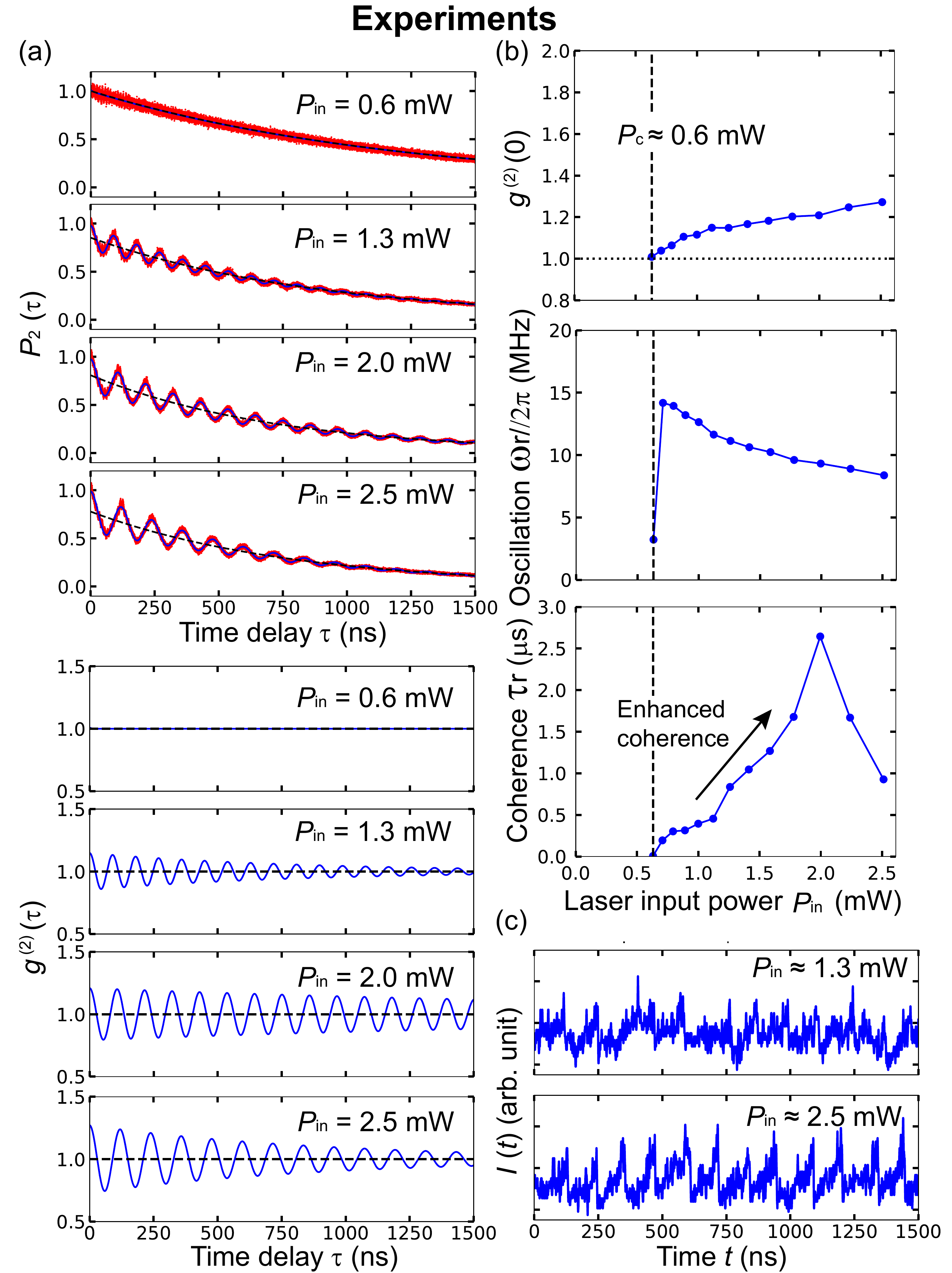}
\caption{(a) Examples of measured photon correlations with the start-stop HBT interferometer $P_2(\tau)$ with a fitting curve by Eq. (\ref{eq:deconv}) and reconstructed normalized second-order photon correlation functions $g^{(2)}(\tau)$. (b) $g^{(2)}(0)$ (top), the oscillation frequency $\omega_{r}$ (middle), and the coherence time $\tau_r$ (bottom) of measured $g^{(2)}(\tau)$. The critical input power of self-pulsing is $P_c\simeq$0.6 mW. (c) Real-time trajectories of the light output measured with an avalanche photodiode (APD) for two pump powers. For measurements, the detuning was fixed as $\delta\simeq-2$.}
\label{fig:g2}
\end{figure}
 
First, we discuss bistable operation, which is shown in Fig. \ref{fig:schematic}(c). When the detuning is $\delta\simeq-2$, the light output intensity $I_{\rm out}$ exhibits a hysteresis loop in terms of laser input power $P_{\rm in}$. We use a negative detuning ($\delta\simeq-2$) to induce the TO nonlinearity. The hysteresis loop shown in Fig. \ref{fig:schematic}(c) is very noisy, which is probably because we performed a single-shot measurement by ramping the laser input power up and down slowly enough to induce the TO nonlinearity. The lower and upper thresholds of the bistable hysteresis loop are about $P_{\rm in}=0.08$ and 0.5 mW, respectively. Note that the laser input power $P_{\rm in}$ was measured as the fiber output of the tunable semiconductor laser. If coupling loss from the fiber output and to the input waveguide is assumed to be 10 dB, the lower threshold power of bistability is 8 $\mu$W in the input waveguide, which is as low as that reported in our previous experiments \cite{Notomi2005,Tanabe2007a}. Thus, the bistable operation in Fig. \ref{fig:schematic}(c) is evidence of the onset of a high-Q cavity-enhanced optical nonlinearity induced by a very small input power. In this paper, the detailed shape of the hysteresis loop is not important, but the separation between the bistable and the self-pulsing region is important for observing the onset of self-pulsing, which was realized by the high Q value of our cavity. This point is covered in more detail in the discussion of Fig. 3(a).
 
Second, for various laser input powers $P_{\rm in}$, we measured the delay-dependent photon correlations $P_2(\tau)$ with the start-stop HBT interferometer, and attempted to reconstruct the normalized second-order photon correlation $g^{(2)}(\tau)$ from $P_2(\tau)$. $P_2(\tau)$ is a histogram of detected photon pairs  in terms of the time delay $\tau$. The upper part of Fig. \ref{fig:g2}(a) shows $P_2(\tau)$ for four laser input powers. The overall exponential decay of the measured $P_2(\tau)$ is a well-known artifact associated with the start-stop measurement \cite{Mandel1995}. Namely, when $\tau$ is longer, the probability of detecting photon pairs becomes smaller. Now, we define $g^{(2)}(\tau)$ as a classical intensity correlation $g^{(2)}(\tau)\equiv\langle{I}(t){I}(t+\tau)\rangle/\langle{I}\rangle^2$, where the brackets represent statistical averages. For reconstructing normalized second-order photon correlation functions $g^{(2)}(\tau)$, we fit the measured $P_2(\tau)$ as
\begin{eqnarray}
P_2(\tau)\simeq C\left[1+A e^{-\frac{|\tau|}{\tau_r}}\cos(\omega_r|\tau|)\right]e^{-\frac{\tau}{\tau_{\rm cor}}},
\label{eq:deconv}
\end{eqnarray}
where $A\equiv g^{(2)}(0)-1$, and $C$ is another fitting parameter. Additionally, $\omega_r$ and $\tau_r$ are the oscillation frequency and coherence (damping) time of $g^{(2)}(\tau)$, respectively. On the other hand, $\tau_{\rm cor}$ is the overall decay time of $P_2(\tau)$ associated with the start-stop counting method. With this fitting, we reconstruct $g^{(2)}(\tau)$ as $g^{(2)}(\tau)=g^{(2)}(0)\cos(\omega_r|\tau|)e^{-{\tau}/{\tau_r}}$. In the lower part of Fig. \ref{fig:g2}(a), we show four reconstructed $g^{(2)}(\tau)$’s corresponding to the four $P_2(\tau)$’s. When laser input power is below a critical value, and even when it is in the bistable hysteresis loop, the light output has a Poissonian fluctuation, and thus $g^{(2)}(\tau)=1$ as shown in Fig. \ref{fig:g2}(a) for $P_{\rm in}=0.6$ mW. Meanwhile, when laser input power is above the critical value, $g^{(2)}(0)$ deviates from unity and $g^{(2)}(\tau)$ exhibits damped oscillation [see $P_{\rm in}=1.3$ mW in Fig. \ref{fig:g2}(a)]. The critical laser input power of damped oscillation of $g^{(2)}(\tau)$ was measured as $P_c\simeq0.6$ mW, which is above the hysteresis loop as shown in Fig. \ref{fig:schematic}(c). In Fig. \ref{fig:g2}(b), we plot the second-order photon correlation at a zero delay time $g^{(2)}(0)$ (top), the oscillation frequency $\omega_r$ (middle), and the coherence time $\tau_r$ (bottom) of $g^{(2)}(\tau)$. Figure \ref{fig:g2}(b) clearly shows that $g^{(2)}(0)$ deviates from unity when $P_{\rm in}=P_c$. Additionally, the oscillation frequency $\omega_r$ has a maximum ($\omega_r/2\pi=14$ MHz) when $P_{\rm in}=P_c$, and it gradually decreases with an increase in laser input power. Meanwhile, for the coherence time $\tau_r$, above $P_c$, $\tau_r$ increases with increasing  laser input power (indicated by an arrow) and reaches a maximum value of $2.6$ $\mu$s when $P_{\rm in}\simeq2.0$. However, when laser input power is increased further, the coherence time $\tau_r$ starts to decrease. The technical details of the hysteresis and $g^{(2)}(\tau)$ measurements are described in Section IC in the SM. 
 
We attribute the origin of the oscillation of $g^{(2)}(\tau)$ to self-pulsation (a limit cycle) originating from Hopf bifurcation \cite{Priem2005,Yacomotti2006,Johnson2006,Pernice2010,Brunstein2012,Cazier2013}. To confirm this, we performed real-time measurements of the light output. Figure \ref{fig:g2}(c) shows real-time trajectories of light outputs measured with the APD for two input powers above $P_c$, which clearly indicates real-time self-pulsation. Thus, here the origin of photon bunching [$g^{(2)}(0)>1$] is the real-time modulation of light intensity \cite{Loudon1980}, which is different from the photon bunching mechanism of chaotic light. Although we performed real-time measurements just to confirm limit cycle oscillation, in principle, we can calculate a classical $g^{(2)}(\tau)$ from the evolution of the light output. This alternative $g^{(2)}(\tau)$ measurement technique is discussed in Section IC in the SM. 
 
Now, a new question arises: What is the origin of the finite coherence time of the observed $g^{(2)}(\tau)$? The answer is the stochastic (noisy) nature of our limit cycle. In fact, without any noise, $g^{(2)}(\tau)$ will never decay and should have an infinite coherence time. For a deeper understanding of these experimental results, we performed numerical simulations based on the coupled-mode equations proposed in Ref. \cite{Rossi2009,VanVaerenbergh2012,Zhang2013}. With the Kerr effects neglected, the normalized coupled-mode equations for an electric field $\alpha$, normalized carrier density $n$, and thermal effect $\theta$ are given by
\begin{eqnarray}
\dot{\alpha}&=&\kappa\lbrace i(-\delta-\theta+n)-(1+fn)\rbrace\alpha+\kappa\sqrt{P_{\rm in}}\label{eq:coupled_alpha}\\
\dot{n}&=&\gamma_n\lbrace-n+\xi|\alpha|^4\rbrace\label{eq:coupled_n}\\
\dot{\theta}&=&\gamma_\theta\lbrace-\theta+\beta|\alpha|^2+\eta|\alpha|^2n\rbrace.\label{eq:coupled_theta}
\end{eqnarray}
Here, $\theta$ is proportional to the temperature difference between the cavity and the surrounding region \cite{VanVaerenbergh2012}. Both $n$ and $\theta$ are normalized to make constants of nonlinear energy shifts in Eq. (\ref{eq:coupled_alpha}) unity. The $\kappa$, $\gamma_n$, and $\gamma_\theta$ are decay rates of the electric field, carrier, and thermal effect, respectively. The field decay rate $\kappa$ includes losses to the waveguides. $P_{\rm in}$ represents normalized laser input power. The coefficients $f$, $\xi$, $\beta$, and $\eta$ represent nonlinear effects associated with free-carrier absorption (FCA), two-photon absorption (TPA), heating with linear photon absorption, and FCA-induced heating, respectively. For these nonlinear coefficients, we use the same values as in Ref. \cite{Zhang2013}: $f=0.0244$, $\xi=8.2\kappa/\gamma_n$, $\beta=0.0296\kappa/\gamma_\theta$, and $\eta=0.0036\kappa/\gamma_\theta$, where the value of $\kappa$ was estimated from the measured $Q$ value. These nonlinear coefficients and their definitions are summarized on Tables S1 and S2 in the SM. For the photon, carrier, and thermal lifetimes, we use $1/2\kappa=300$ ps, $1/\gamma_n=200$ ps, and $1/\gamma_\theta=100$ ns, respectively. The fast carrier lifetime ($1/\gamma_n=200$ ps) results from fast carrier diffusion associated with the small cavity of the PhC structure \cite{Tanabe2005,Tanabe2008}.
 
\begin{figure}
\includegraphics[width=0.48\textwidth]{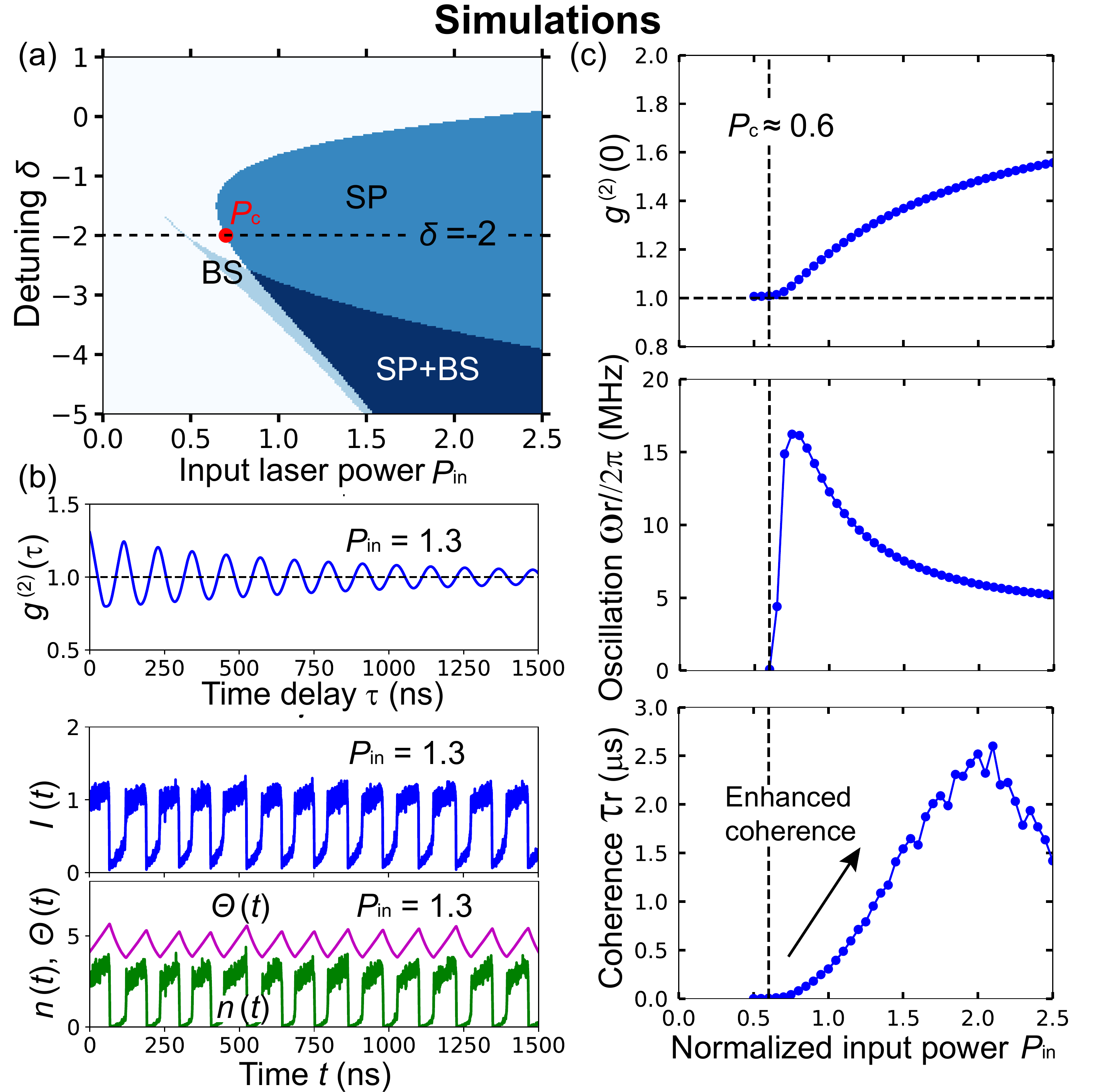}
\caption{(a) Simulated self-pulsing (SP) and bistable (BS) regions as a map of detuning $\delta$ and laser input power $P_{\rm in}$, where the red filled circle represents the onset of SP for $\delta=-2$. (b) Simulated $g^{(2)}(\tau)$ and real-time evolution of carrier $n(t)$, thermal effect $\theta(t)$, and the light output $I(t)=|\alpha(t)|^2$ for $P_{\rm in}=1.3$. (c) $g^{(2)}(0)$ (top), oscillation frequencies $\omega_{r}$ (middle), and the coherence time $\tau_r$ (bottom) of the simulated $g^{(2)}(\tau)$. For simulations, the detuning $\delta=-2$ was used. The critical input power of self-pulsing is $P_c$=0.6. While (a) is the result with the deterministic coupled-mode equations, (b) and (c) are the results with stochastic coupled-mode equations with noise terms.}
\label{fig:sim}
\end{figure}

Before showing the simulations of stochastic dynamics, we briefly investigate the static properties of the deterministic coupled-mode Eqs. (\ref{eq:coupled_alpha})-(\ref{eq:coupled_theta}). First, we attempt to obtain steady state values of $\alpha$, $n$, and $\theta$, which are denoted as $\alpha_{s}$, $n_{s}$, and $\theta_{s}$, respectively. By putting $\dot{\alpha}=0$, $\dot{n}=0$, and $\dot{\theta}=0$ into Eqs. (\ref{eq:coupled_alpha})-(\ref{eq:coupled_theta}), we obtain an algebraic equation for $I_{s}=|\alpha_{s}|^2$ (see Section IIB in the SM for the explicit form of the algebraic equation). The system has two equilibria when the algebraic equation has two solutions for $I_{s}$. Second, at the steady state values of $\alpha_{s}$, $n_{s}$, and $\theta_{s}$, we calculate a Jacobian matrix and its eigenvalues to find self-pulsing (see Section IIB in the SM for the explicit form of the Jacobian and their eigenvalues). When a pair of the eigenvalues have positive real parts, the dynamical system becomes unstable, and Hopf bifurcation (self-pulsing) occurs \cite{Strogatz2018,Kuramoto2003}. Our system has the following three regions: a self-pulsing (SP) region where a single unstable equilibrium exists, a bistable (BS) region where there are two stable equilibria, and an SP+BS region where one equilibrium is stable and the other is not. The diagram of our dynamical system is shown in Fig. \ref{fig:sim}(a). We also found that bistability is induced solely by TO nonlinearity, while self-pulsing requires both carrier and TO nonlinearities (see Section IIC in the SM). The horizontal dashed line in Fig. \ref{fig:sim}(a) indicates that, for $\delta=-2$, with an increase in pump power, self-pulsing occurs when input power reaches a critical power $P_c$, which is larger than the upper threshold of the bistable hysteresis loop. This is consistent with our measurement shown in Fig. \ref{fig:schematic}(c). We comment on the importance of the separation between the SP and BS regions shown in in Fig. \ref{fig:schematic}(c). In our experiment, we were able to observe the onset (bifurcation point) of self-pulsing outside the hysteresis loop with moderate negative detuning ($\delta\simeq-2$) and low input power ($P_c\simeq0.6$ mW). We found that as photon lifetime increases (a $Q$ value increases), the SP region separates from the BS region, and self-pulsing occurs with a near-zero detuning and low input power. Thus, a high $Q$ value is technically very important for observation of the onset of self-pulsing. Further discussion on the impact of $Q$ on self-pulsing is given in Section III in the SM, where simulations for a moderate $Q=2.0\times10^4$ value are shown. Even in the moderate $Q$ cavity, non-trivial regions are only the SP, BS, and SP+BS regions in the same way as in Fig. \ref{fig:sim}(a). However, the shapes of these regions as functions of $\delta$ and $P$ are very different from those in Fig. \ref{fig:sim}(a). 
 
Now, we investigate the dynamical properties of coupled-mode Eqs. (\ref{eq:coupled_alpha})-(\ref{eq:coupled_theta}). Since we are interested in a fluctuating system, we add additive Langevin noises $f_x$ and $f_y$ only to Eq. (\ref{eq:coupled_theta}), where we assume that field and laser input noises are dominant over other noises. Actually, we find that the inclusion of carrier and thermal noises does not qualitatively modify the results. The noise terms satisfy correlations $\langle f_i(t)f_j(t')\rangle=2D_\alpha\delta_{i,j}\delta(t-t')$ and $\langle f_i(t)\rangle=0$, where $i(j)=x,y$ and the coefficient $D_\alpha$ is the strength of the noise. For numerical simulations of the stochastic equations, we employed the Euler-Maruyama method. The value of the noise strength was set as $\sqrt{2D_\alpha}=0.05\sqrt{\kappa}$, which was chosen to reproduce the observed maximum coherence time of $g^{(2)}(\tau)$ \footnote{In actual numerical simulations, we introduce the noise to a difference equation as $\sigma\xi_R\sqrt{dt}$, where $\sigma=0.05\sqrt{\kappa}$ and $\xi_R$ is a random umber following the normal distribution $N(0,1)$.}. Figure \ref{fig:sim}(b) shows simulated $g^{(2)}(\tau)$, $n(t)$, $\theta(t)$, and $I(t)=|\alpha(t)|^2$ for input power $P_{\rm in}=1.3$, which clearly reproduce the damped oscillatory behavior of $g^{(2)}(\tau)$ and the real-time self-pulsing when the input power is above the critical input power $P_c=0.6$. Additionally, in Fig. \ref{fig:sim}(c), we plot $g^{(2)}(0)$ (top), the oscillation frequency $\omega_r$ (middle), and the coherence time $\tau_r$ (bottom) of $g^{(2)}(\tau)$ as a function of $P_{\rm in}$, which also qualitatively reproduce the measurements shown in Fig. \ref{fig:g2}(b). Namely, the simulation reproduces the monotonic decrease of $\omega_r$ and the enhancement and reduction of the coherence time $\tau_r$ with an increase in pump power. Here, we briefly comment on the reduction of the coherence time $\tau_r$ in the high input power region ($P_{\rm in}>2.0$ mW in the experiment). We found that field and carrier noises give rise to the coherence reduction, while thermal noise does not. Simulations with carrier and thermal noises are shown in Section IV in the SM, which indicates that the thermal noise may be negligible.

In the rest of this paper, we focus on the region around the critical input power of self-pulsing and attempt to interpret the enhancement of the coherence time $\tau_r$, which is indicated by an arrow in the bottom graphs of Fig. \ref{fig:g2}(b) (experiment) and Fig. \ref{fig:sim}(c) (simulation). For this purpose, we employ the phase reduction theory, which starts from defining a generalized phase, $\phi$, along a limit cycle’s orbit. Importantly, in phase reduction, noises in a limit cycle are reduced to a frequency drift and a phase noise as $\dot{\phi}=\omega+v+f_\phi$, where $\omega$, $v$, and $f_\phi$ represent the original frequency of a limit cycle, the frequency drift, and the phase noise, respectively \cite{Kuramoto2003,Nakao2017}. The phase noise $f_\phi$ satisfies correlations $\langle f_\phi(t)f_\phi(t')\rangle=2D_\phi\delta(t-t')$ and $\langle f_\phi(t)\rangle=0$, where $D_\phi$ is the phase diffusion rate. Therefore, for a limit cycle, noises are interpreted as diffusion of the generalized phase. Furthermore, the coherence time of a correlation function such as $g^{(2)}(\tau)$ corresponds to the coherence (diffusion) time of the generalized phase: $\tau_r\simeq1/D_\phi$ \cite{Cao2015,Fei2018}. 
 
To be more concrete, let us recall that for the stochastic Stuart-Landau model without phase-amplitude coupling, the phase diffusion rate well above Hopf bifurcation is approximated as \cite{Louisell1973,VanKampen1992,Risken1996,Cao2015,Fei2018}
\begin{equation}
D_\phi\propto D_0/P_{\rm in},\label{eq:diffusion}
\end{equation}
where $D_0$ is the strength of noises, while $P_{\rm in}$ represents the pump or input power to the system. If $D_0$ is constant, Eq. (\ref{eq:diffusion}) represents suppression of phase diffusion by pumping \footnote{One may find that Eq. (\ref{eq:diffusion}) is analogous to the well-known Schawlow-Townes linewidth reduction in laser physics.}. Additionally, Ref. \cite{Cao2015} shows that Eq. (\ref{eq:diffusion}) can also be written with a free-energy dissipation rate $\Delta W$ as $D_\phi\propto\Delta W^{-1}$, which means suppression of phase diffusion through free-energy dissipation. Furtheremore, Eq. (\ref{eq:diffusion}) is intuitively understood as a one-dimensional diffusion process along an orbit of a limit cycle \cite{Scully1999}. Thus, if the noise strength $D_0$ is constant, the longer the orbit's circumference, the longer the time required for the phase to diffuse over $2\pi$. Additionally, the amplitude and the circumference length generally increase with pumping $P_{\rm in}$ in the vicinity of Hopf bifurcation. Thus, the essence of Eq. (\ref{eq:diffusion}) lies in the fixed strength of noises and the increase in the amplitude by pumping. In particular, the latter is possible only for limit cycles. Thus, in the vicinity of a critical point of self-pulsing, an enhancement of phase coherence will generally occur for any limit cycle, including ours. 

In summary, we performed photon correlation measurements of stochastic limit cycles using a driven high-$Q$ silicon photonic crystal cavity. We observed damped oscillation of photon correlation associated with self-pulsing (a limit cycle). Furthermore, by increasing input power, the coherence time of the photon correlation function was enhanced up to the order of microseconds, which could be interpreted as coherence time enhancement of a generalized phase through pumping. 
 
\textit{Note}. During preparation of the manuscript, we noticed a paper with similar keywords \cite{Marconi2019} 

\section*{Acknowledgements}
We thank K. Nozaki for helpful discussions. 

\clearpage

\section{Experimental details}
\begin{figure}
\includegraphics[width=0.48\textwidth]{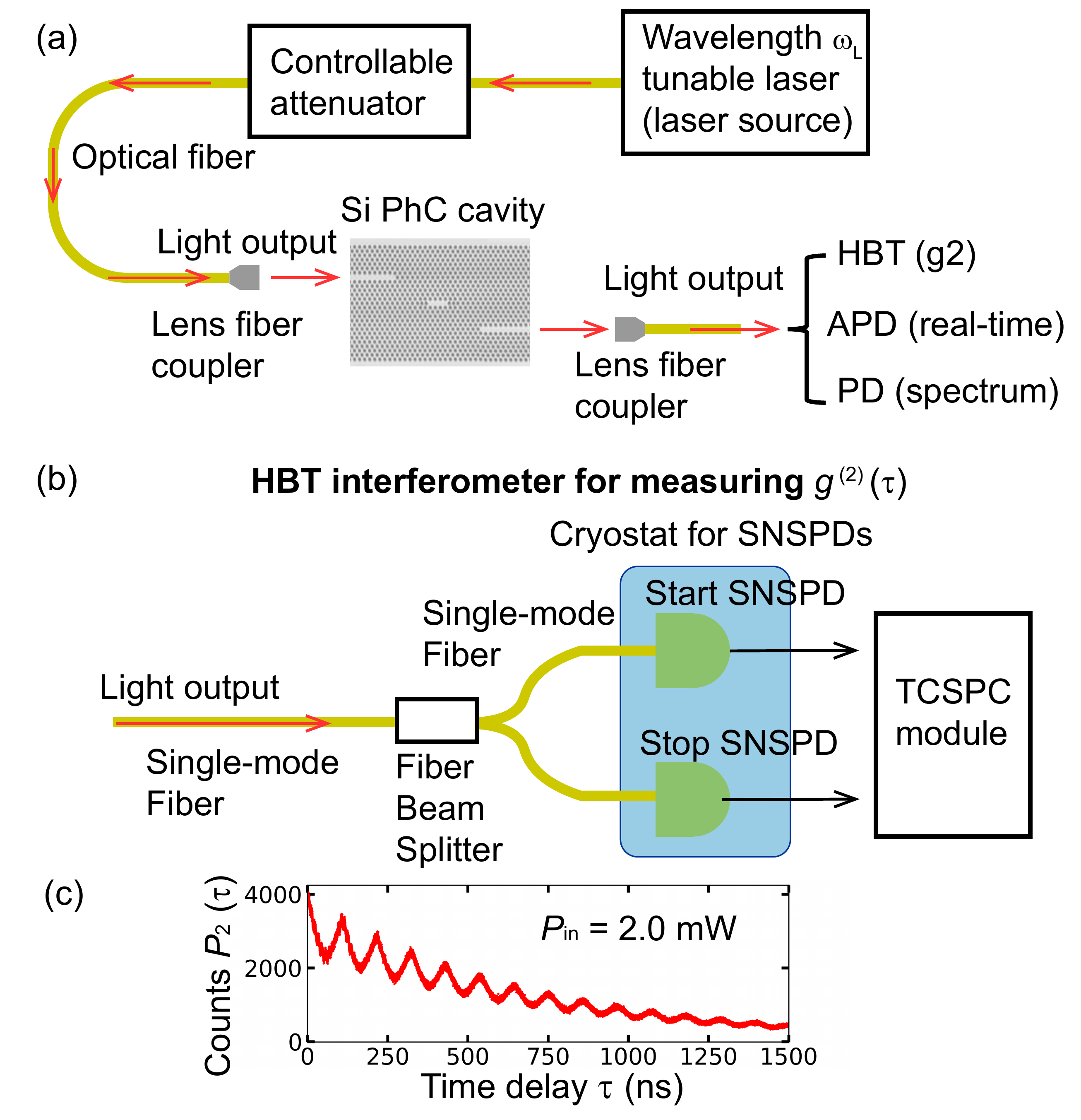}
\caption{(a)  Schematic of optical setup for measuring resonance spectrum, second-order photon correlation functions, and real-time evolutions. (b) Start-stop Hanbury brown-Twiss (HBT) interferometer is illustrated, which is composed of a fiber beam splitter and two superconducting nanowire single-photon detectors (SNSPDs). TCSPC stands for time-correlated single photon counting. (c) Histogram $P_2(\tau)$ obtained with start-stop counting for $P_{in}=2.0$.}
\label{fig:Supple_setup}
\end{figure}

We describe experimental details. The general optical setup for our experiments is depicted in Fig. \ref{fig:Supple_setup}(a). Using two lens fiber couplers, the laser input is coupled to the input waveguide, and the light output is collected from the output waveguide. The laser input originates from a semiconductor laser source whose wavelength is tunable between 1460 and 1640 nm. In the experiments, the collected light output was carried to  the measurement devices with an optical fiber: the HBT interferometer, avalanche photodiode (APD), and photodiode. Laser input power was adjusted by the power of the laser source and controllable attenuator.

\subsection{Silicon photonic crystal microcavity}
The detailed structure and fabrication method of our silicon (Si) photonic crystal (PhC) cavity have already been reported in Ref. \cite{Kuramochi2014}. The PhC structure was made on Si wafer with positive tone resists and by inductively coupled plasma (ICP) etching. After this process, an air-bridge PhC membrane structure was created using buffered HF treatment. As described in the main text, the PhC slab is a two-dimensional hexagonal lattice with a lattice constant 412 nm, air-hole radius 100 nm, and thickness of 215 nm. The optical cavity is based on the L3-like structure, where three air-holes are removed. In addition to the three removed air-holes, for a further increase in the $Q$ value, the positions of several air-holes around the cavity region are carefully modulated. The details of this modulation of air-hole positions are described as ``Type III" in Ref. \cite{Kuramochi2014}. For coupling between the cavity and the waveguides, we used a $\Gamma$-M coupling configuration. In regard to fabrication errors, the accuracy of the air-holes is better than 1 nm, while the standard deviation of air-hole radii is about 1 nm. The shift of the resonance wavelength and variation of $Q$ value originate mainly from the fluctuation of air-hole radii. 

\subsection{Resonance spectrum and hysteresis curve}
First, we estimated the cavity's $Q$ value from the Lorentzian fitting of the resonance spectrum shown in Fig. 1(d) in the main text. The $Q$ value is defined as $Q\equiv\Delta\lambda/\lambda$, where $\Delta\lambda$ is the full width at half maximum (FWHM) of the spectrum, while $\lambda$ is the center wavelength of the resonance. From the measured $Q$ value, the cavity photon lifetime $1/2\kappa$ was calculated as $\frac{1}{2\kappa}=\frac{Q\lambda}{2\pi c}$. The resonance spectrum was obtained by measuring light output intensities as a function of the wavelengths of the tunable laser. For measuring the spectrum of the cavity, the laser input power was fixed to 0.01 mW, which was below the hysteresis loop and thus low enough to avoid nonlinearities. Also note that the measured $Q$ value was the ``total $Q$ value" that includes coupling losses to the waveguides. 

Second, to obtain the hysteresis loop in Fig. 1(c) in the main text, we temporally modulated the attenuation of the controllable attenuator, while the power of the laser source was fixed. The sweep speed of the controllable attenuator was 0.25 dB/s. The time required for the one-way sweep (from 0 to 20 dB) was around 80 s. Additionally, for each step, the measurement time for the light output intensity was 0.8 s. All these time scales are much slower than the thermal dynamics ($\sim$100 ns), which are the slowest dynamics in the system. 

\subsection{Photon correlation measurement}
We describe the details of photon correlation measurements. As depicted in Fig. \ref{fig:Supple_setup}(b), photon correlation functions $g^{(2)}(\tau)$ were measured with the start-stop HBT interferometer. For detection, we used two superconducting nanowire single-photon detectors (SNSPDs). More precisely, we use the two channels of a SNSPD unit. The advantage of using SNSPDs lies in their high quantum efficiencies (more than 80\%) and extremely low dark count rate (less than 10 counts/s), which leads to a high signal-to-noise ratio (SNR). Although the dead time of an SNSPD is relatively long ($\sim10$ ns), we are able to overcome this problem by employing the HBT interferometer, which is composed of a fiber beam splitter and two SNSPDs for ``start" and ``stop" photon counting. In our measurements with the HBT interferometer, time-resolution is actually limited by a 30-ps timing jitter.

We explain the start-stop photon counting for constructing second-order photon correlation functions. We recorded a histogram $P_2(\tau)$ of detected photon pairs in terms of the time delay $\tau$ [see Fig. \ref{fig:Supple_setup}(c)]. In more detail, after the ``start" SNSPD's count of a photon, we recorded the time delay $\tau$ required for the ``stop" SNSPD to detect a photon. It is more probable that the stop SNSPD  detects a photon with a shorter delay time, which results in the overall exponential decay of $P_2(\tau)$ shown in Fig. \ref{fig:Supple_setup}(c). The time bin for $\tau$ was 0.0160 ns, and the counting number for the zero-delay ($\tau=0$) time bin was typically 3,000-4,000   counts. 

\subsection{Real-time measurement}
For real-time measurements, we used a linear-mode APD. Note that our APD does not have a single-photon counting mode (a Geiger mode). The APD used for real-time measurements has a bandwidth of $\sim$1 GHz, and thus the time resolution is about 1 ns. The noise-equivalent power (NEP) of the APD is 1.6 pW/$\sqrt{\rm Hz}$. This time resolution was short enough to measure the time evolution of self-pulsing.

\begin{figure}
\includegraphics[width=0.48\textwidth]{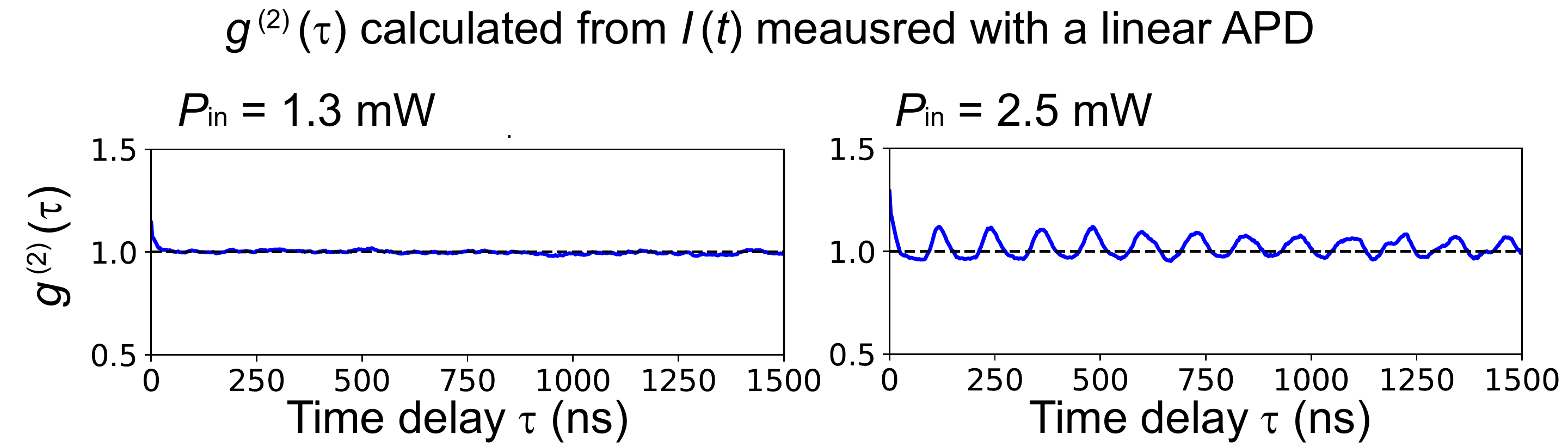}
\caption{Classical second-order correlations $g^{(2)}(\tau)$ calculated from single-shot time evolutions obtained with a linear-mode APD. The two $g^{(2)}(\tau)$ correspond to the time evolutions $I(t)$ for $P_{\rm in}=1.3$ and 2.5 mW shown in Fig. 2(c) in the main text. To calculate $g^{(2)}(\tau)$, we used time evolutions until 5,000 ns. Meanwhile, in Fig. 2(c) in the main, we showed time evolutions only until 1,500 ns text.}
\label{fig:g2_APD}
\end{figure}

We also comment on an alternative method to obtain second-order correlation functions $g^{(2)}(\tau)$ using a linear-mode APD \cite{Wang2015}. In principle, a classical correlation $g^{(2)}(\tau)\equiv\langle{I}(t){I}(t+\tau)\rangle/\langle{I}\rangle^2$ can be calculated from time evolution $I(t)$ obtained by an APD [see Fig. 2(c) in the main text]. In Fig. \ref{fig:g2_APD}, we show two classical correlations $g^{(2)}(\tau)$ calculated from the time evolutions $I(t)$ for $P_{\rm in}=1.3$ and 2.5 mW shown in Fig. 2(c) in the main text. In fact, as Fig. \ref{fig:g2_APD} shows, this technique is attractive because the classical $g^{(2)}(\tau)$ does not have the overall decay associated with the start-stop counting. However, we employed the start-stop HBT interferometer and single-photon counting detectors (SNSPDs), mainly because of the small dark count rate of single-photon counting devices compared with a linear-mode APD. The small dark count rate is very important for high-SNR photon statistical measurements of low intensity light. Moreover, it is difficult to determine ``zero intensity" with a linear-mode APD because a photocurrent signal is always present even without light inputs. Of course, $g^{(2)}(\tau)$ with a high SNR can also be obtained with single-photon-counting-mode (Geiger-mode) APDs and the HBT interferometer. Actually, Fig. \ref{fig:g2_APD} indicates that for the strong signal ($P_{\rm in}=2.5$ mW), the classical correlation has a damped oscillatory behavior similar to that in Fig. 2(a) in the main text, while for the weak signal ($P_{\rm in}=1.3$ mW), we could not obtain a significant correlation due to the low SNR of the linear-mode APD. 

Finally, for future perspectives, we discuss the best $g^{(2)}(\tau)$ measurement technique. Let us recall that the overall decay of $P_2(\tau)$ shown in Fig. 2(a) in the main text originated from the start-stop counting. If we record the times of single-photon ``clicks" by the two SNSPDs, a numerically reconstructed photon correlation function $g^{(2)}(\tau)$ does not have the overall decay. This method is similar to $g^{(2)}(\tau)$ measured by a photon-counting streak-camera \cite{Wiersig2009,Takemura2012}, which records the history of photon arrivals. In summary, in the best photon correlation measurement technique, $g^{(2)}(\tau)$ obtained from the histories of single-photon clicks achieves a high SNR and does not have the overall decay.

\section{Theoretical model}
We describe the details of the theoretical model mentioned in the main text. In particular, we detail the coupled-mode equations, linear stability analysis, and the effects of the carrier- and thermo-optic-induced nonlinearities. 

\subsection{Coupled-mode equations}
First, coupled-mode Eqs. (2)-(4) in the main text were derived from the more complete and normalized model proposed in Ref. \cite{Zhang2013}:
\begin{eqnarray}
\dot{\alpha}&=&\kappa\left[ i(-\delta-n_{\rm Kerr}|\alpha|^2-\theta+N+\sigma_{\rm FCD}N^{0.8})\right.\nonumber\\
&&\left.-(1+\alpha_{\rm TPA}|\alpha|^2+\gamma_{\rm FCA}N)\right]\alpha+\kappa\sqrt{P_{\rm in}}\label{eq:a_Zhang}\\
\dot{N}&=&\kappa\left[-(\gamma_n/\kappa)N+|\alpha|^4\right]\label{eq:n_Zhang}\\
\dot{\theta}&=&\kappa\left[-\gamma_\theta\theta+\xi_T\eta_{\rm lin}|\alpha|^2+2\xi_T\gamma_{\rm FCA}|\alpha|^2N\right.\nonumber\\
&&\left.+2\alpha_{\rm TPA}\xi_T|\alpha|^4\right].\label{eq:t_Zhang}
\end{eqnarray}
Here, $N$ represents the carrier density, and is proportional to $n$ in the main text. For all Si nonlinear coefficients in Eqs. (\ref{eq:a_Zhang})-(\ref{eq:a_Zhang}), we used the same values as in Ref. \cite{Zhang2013}. Table \ref{table1} summarizes the physical meanings and values of the symbols. 
\begin{table}[htbp]
\caption{Parameter values used in Eqs. (\ref{eq:a_Zhang})-(\ref{eq:t_Zhang})}
 \centering
 \begin{tabular}{l c r r}
 \hline\hline
 Symbol & Description & Value \\ 
 \hline
 $n_{\rm Kerr}$ & Kerr effect & 0.55\\
 $\sigma_{\rm FCD}$ & Free-carrier dispersion (FCD) effect & 7.2\\ 
 $\alpha_{\rm TPA}$ & Two-photon absorption (TPA) effect & 0.11\\
 $\gamma_{\rm FCA}$ & Free-carrier absorption (FCA) effect & 0.20\\
 $\xi_T$ & & 0.074\\
 $\eta_{\rm lin}$ & Fraction of linear loss due to absorption \cite{Zhang2013,VanVaerenbergh2012}& 0.4\\
 \hline\hline
 \end{tabular}
 \label{table1}
\end{table}
To obtain the coupled-mode equations used in the main text, we performed further simplification following Ref. \cite{VanVaerenbergh2012}. First, in Eq. (\ref{eq:a_Zhang}), we approximated the FCD related term as
\begin{equation}
N+\sigma_{\rm FCD}N^{0.8}\simeq g_nN\ \ {\rm with}\ \ g_n\equiv1+\sigma_{\rm FCD}.
\end{equation} 
Second, we neglected the Kerr $n_{\rm Kerr}|\alpha|^2$ and TPA effects $\alpha_{\rm TPA}|\alpha|^2$ in Eq. (\ref{eq:a_Zhang}). We also neglected the TPA effect $2\alpha_{\rm TPA}\xi_T|\alpha|^4$ in Eq. (\ref{eq:t_Zhang}). We neglected the Kerr and TPA related terms because the cavity's field intensity in our experiments is sufficiently small for the onsets of these effects. Third, we introduced a new variable $n$, which is proportional to $N$ as 
\begin{equation}
n\equiv g_nN.
\end{equation}
With these approximations and the introduction of the variable $n$, Eqs. (\ref{eq:a_Zhang})-(\ref{eq:a_Zhang}) become
\begin{eqnarray}
\dot{\alpha}&=&\kappa\left[ i(-\delta-\theta+n)-\left(1+\frac{\gamma_{\rm FCA}}{g_n}n\right)\right]\alpha+\kappa\sqrt{P_{\rm in}}\nonumber\\
&&\label{eq:a_main}\\
\dot{n}&=&\gamma_n\left[-n+g_n\frac{\kappa}{\gamma_n}|\alpha|^4\right]\label{eq:n_main}\\
\dot{\theta}&=&\gamma_\theta\left[-\theta+\xi_T\eta_{\rm lin}\frac{\kappa}{\gamma_\theta}|\alpha|^2+\frac{2\xi_T\gamma_{\rm FCA}}{g_n}\frac{\kappa}{\gamma_\theta}|\alpha|^2n\right],\label{eq:t_main}
\end{eqnarray}
which are coupled-mode Eqs. (2)-(4) in the main text if we introduce coefficients defined as $f= \gamma_{\rm FCA}/g_n$, $\xi=g_n\kappa/\gamma_n$, $\beta=\xi_T\eta_{\rm lin}\kappa/\gamma_\theta$, and $\eta=({2\xi_T\gamma_{\rm FCA}}/{g_n})\kappa/\gamma_\theta$. The definitions and values of these coefficients are summarized in Table \ref{table2}. 
\begin{table}[htbp]
\caption{Parameter values used in Eqs. (\ref{eq:a_main})-(\ref{eq:t_main})}
 \centering
 \begin{tabular}{l c r r}
 \hline\hline
 Symbol & Definition & Value \\ 
 \hline
 $f$ & $\frac{\gamma_{\rm FCA}}{g_n}$ & 0.0244\\
 $\xi$ & $g_n\frac{\kappa}{\gamma_n}$ & $8.2\kappa/\gamma_n$\\ 
 $\beta$ & $\xi_T\eta_{\rm lin}\frac{\kappa}{\gamma_\theta}$ & $0.0296\kappa/\gamma_\theta$\\
 $\eta$ & $\frac{2\xi_T\gamma_{\rm FCA}}{g_n}\frac{\kappa}{\gamma_\theta}$ & $0.0036\kappa/\gamma_\theta$\\
 \hline\hline
 \end{tabular}
 \label{table2}
\end{table}

Finally, we comment on the difficulty in determining the exact values of the nonlinear coefficients. Since the coupled-mode equations include four parameters and three lifetimes, it is almost impossible to determine the exact values of these parameters. The most important criterion may be qualitative reproduction of the measured results. In our case, a model has to reproduce self-pulsing and bistability. 
\begin{figure}
\includegraphics[width=0.48\textwidth]{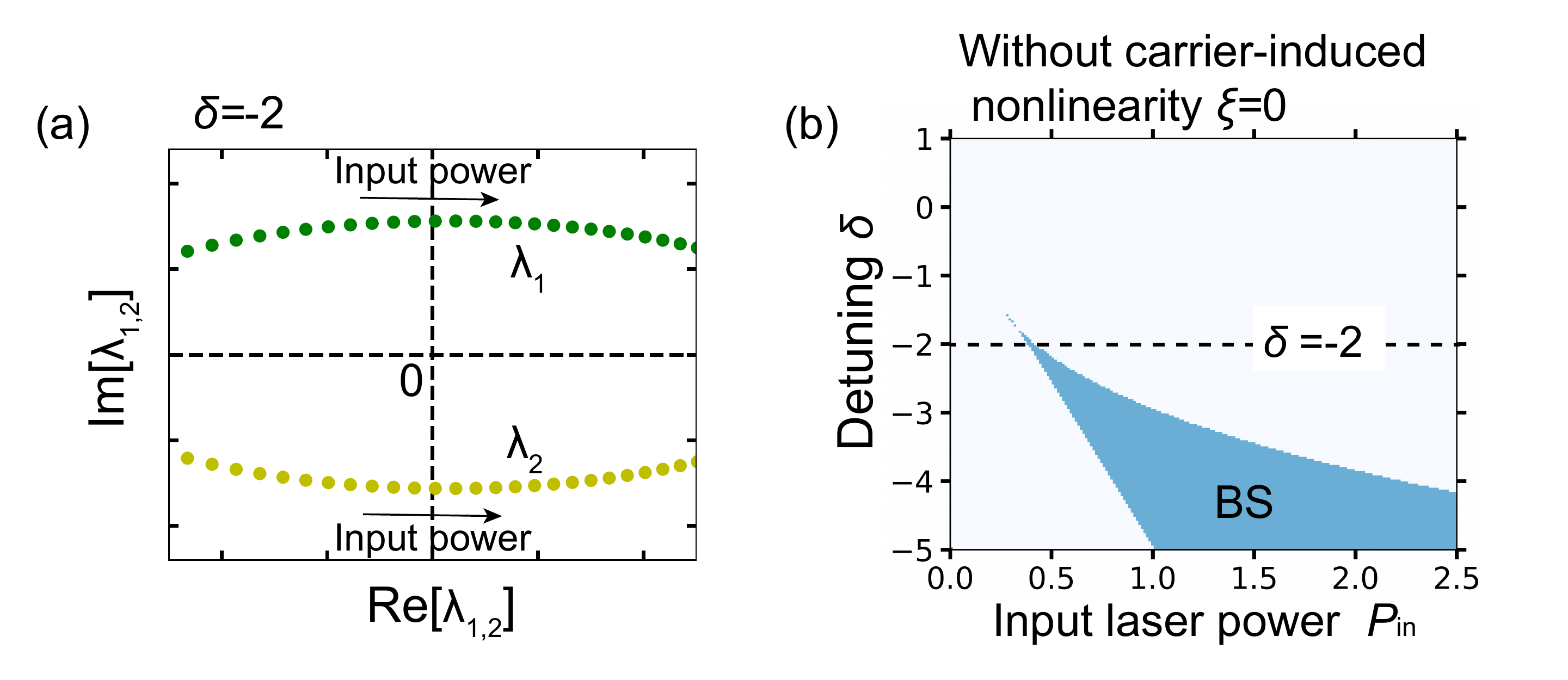}
\caption{(a) Trajectories of two conjugate eigenvalues that cross the imaginary axis when input power $P_{\rm in}$ reaches the critical input power for self-pulsing $P_c$. We show 30 eigenvalues from $P_{\rm in}=0.65$ to $0.75$. (b) The same diagram as that in Fig. 3(a) in the main text, but with only the thermo-optic nonlinearity. To plot it, we set $\xi=0$ to turn off the carrier-induced nonlinearity.}
\label{fig:Supple_eigen}
\end{figure} 

\subsection{Linear stability analysis}
Here, we detail the linear stability analysis. First, we rewrite coupled-mode Eqs. (2)-(4) in the main text in terms of the two components $x=\Re[\alpha]$ and $y=\Im[\alpha]$ of the complex field $\alpha=x+iy$. With $x$ and $y$, the coupled-mode equations are written as
\begin{eqnarray}
\dot{x}=f_x &=& \kappa\lbrace -(1+f n) x-(-\delta-\theta+n) y+\sqrt{P_{\rm in}}\rbrace\nonumber\\
\label{eq:fx}\\
\dot{y}=f_y &=& \kappa\lbrace -(1+f n) y+(-\delta-\theta+n) x\rbrace\\
\dot{n}=f_n &=& \gamma_n\left\lbrace-n+\xi (x^2+y^2)^2\right\rbrace\\
\dot{\theta}=f_\theta &=& \gamma_\theta\left\lbrace -\theta+\beta (x^2+y^2)+\eta (x^2+y^2)n\right\rbrace.
\label{eq:ft}
\end{eqnarray}

First, as briefly mentioned in the main text, we calculate steady state values $\alpha_{s}$, $n_{s}$, and $\theta_{s}$. Putting $\dot{\alpha}=0$, $\dot{n}=0$, and $\dot{\theta}=0$ into Eqs. (\ref{eq:fx})-(\ref{eq:ft}), we obtain an algebraic equation:
\begin{eqnarray}
P_{\rm in}&=&f_{\rm s}(I_{\rm s})\nonumber\\
&=&I_{s}\left[(-\delta-\beta I_{s}-\eta\xi I_{s}^3+\xi I_{s}^2)^2+(1+f\xi I_{s}^2)^2\right].\nonumber\\\label{eq:cubic}
\end{eqnarray}
For various laser input powers $P_{\rm in}$ and detuning values $\delta$, we numerically solved Eq. (\ref{eq:cubic}) and obtained $I_{s}$. Curve $y=f_{s}(x)$ behaves similarly to a cubic function and crosses  curve $y=P_{\rm in}$ at one or two points depending on the values of $P_{\rm in}$ and $\delta$. When Eq. (\ref{eq:cubic}) has two solutions for $I_{s}$, there are two equilibria (two sets of steady states), which we may refer to as high- and low-intensity states. We also note that the two sets of steady states physically belong to the same mode (for instance, the same cavity's optical mode), and only the values are different. Using $I_{s}$, $n_{s}$ and $\theta_{s}$ are easily calculated as 
\begin{eqnarray}
n_{s}=\frac{\xi}{\gamma_n} I_{s}^2\ \ \ {\rm and}\ \ \theta_{\rm s}=\frac{\beta}{\gamma_\theta} I_{s}+\frac{\eta\xi}{\gamma_\theta\gamma_n} I_{s}^3,
\end{eqnarray}
respectively. With $n_{s}$ and $\theta_{s}$, the complex electric field $\alpha_{s}$ is given by
\begin{eqnarray}
\alpha_{s}=\frac{\sqrt{P_{\rm in}}}{\kappa}\cdot\frac{(1+fn_{s})+i(-\delta-\theta_{s}+n_{s})}{(-\delta-\theta_{s}+n_{s})^2+(1+fn_{s})^2}.
\label{eq:ss_alpha}
\end{eqnarray}

Now, using a Jacovian matrix, we investigate the stability of the steady states ($\alpha_{s}$, $n_{s}$, and $\theta_{s}$) obtained above. The $4\times4$ Jacobian matrix ${\bm J}$ corresponding to Eqs (\ref{eq:fx})-(\ref{eq:ft}) is given by
\begin{widetext}
\begin{equation}
{\bm J}
=
\left( \begin{array}{cccc}
\frac{\partial f_x}{\partial x} & \frac{\partial f_x}{\partial y} & \frac{\partial f_x}{\partial n} & \frac{\partial f_x}{\partial \theta}\\
\frac{\partial f_y}{\partial x} & \frac{\partial f_y}{\partial y} & \frac{\partial f_y}{\partial n} & \frac{\partial f_y}{\partial \theta}\\
\frac{\partial f_n}{\partial x} & \frac{\partial f_n}{\partial y} & \frac{\partial f_n}{\partial n} & \frac{\partial f_n}{\partial \theta}\\
\frac{\partial f_\theta}{\partial x} & \frac{\partial f_\theta}{\partial y} & \frac{\partial f_\theta}{\partial n} & \frac{\partial f_\theta}{\partial \theta}
\end{array} \right)
=
\left( \begin{array}{cccc}
\kappa(-fn - 1) & \kappa(\delta - n + t) & \kappa(-fx - y) & \kappa y\\
\kappa(-\delta + n - t) & \kappa(-fn - 1) & \kappa(-fy + x) & -\kappa x\\
4\gamma_n \xi x (x^2 + y^2) & 4\gamma_n\xi y (x^2 + y^2) & -\gamma_n & 0\\
\gamma_\theta(2\beta x + 2\eta n x) & \gamma_\theta(2\beta y + 2\eta n y) & \gamma_\theta\eta (x^2 + y^2) & -\gamma_\theta
\end{array} \right).
\label{eq:Jacobian}
\end{equation}
\end{widetext}
As we have explained in the main text, when a pair of conjugate eigenvalues of the Jacobian matrix ${\bm J}$ have positive real parts, Hopf bifurcation occurs. In Fig. \ref{fig:Supple_eigen}(a), the pairs of the eigenvalues around the bifurcation point ($P_{\rm in}\simeq P_c$) are plotted in a complex plane, where we used $\delta=-2$, and the parameters are the same as those in Fig. 3 in the main text. Figure \ref{fig:Supple_eigen}(a) clearly shows that the pair of the eigenvalues crosses the imaginary axis, which is the onset of Hopf bifurcation \cite{Strogatz2018,Kuramoto2003}. 

Finally, calculating the eigenvalues of the Jacobian at steady state values $\alpha_{s}$, $n_{s}$, and $\theta_{s}$, we obtained the bistability (BS) and self-pulsing (SP) regions as functions of $P_{\rm in}$ and $\delta$, which are shown in Fig. 3(a) in the main text. The inside and outside of the SP region are also interpreted as unstable and stable regions, respectively. Here, we comment on the SP+BS region shown in Fig. 3(a) in the main text. In the SP+BS region, there are two equilibria, but the high-intensity steady-state is unstable, while the low-intensity steady-state is stable.
 
\subsection{Carrier- and thermo-optic-induced nonlinearity}
Here, we discuss the origin of bistability and self-pulsing by turning off the carrier-induced or thermo-optic- (TO) induced nonlinearity. First, we show a diagram only with the TO-induced nonlinearity in Fig. \ref{fig:Supple_eigen}(b), where we set $\xi=0$, and the other parameters are the same as those in Fig. 3(a) in the main text. Since $\xi=0$, carriers are not generated, and no carrier-induced nonlinearity is induced. In other words, in Fig. \ref{fig:Supple_eigen}(b), only the TO-induced nonlinearity is present. Interestingly, in Fig. \ref{fig:Supple_eigen}(b), there is no SP region, while the BS region is completely the same as in Fig. 3(a) in the main text. Second, we are interested in the case where solely carrier-induced nonlinearity is present. However, we are not able to show a corresponding diagram, because neither BS nor SP regions appear when the TO-induced nonlinearity is turned off ($\beta=\eta=0$). Therefore, as in the main text, we conclude that BS is induced solely by the TO nonlinearity, while SP requires both carrier- and TO-induced nonlinearities.

\section{Simulation for moderate $Q$ value}
We discuss the bistable behavior and self-pulsing of a Si PhC microcavity with a moderate $Q$ value. For instance, let us consider a conventional L3 cavity without careful modulation of air-holes, where the photon lifetime is much shorter than the carrier and thermal lifetimes:
\begin{equation}
\kappa\gg\gamma_n,\gamma_\theta.\label{eq:ad_condition}
\end{equation}
Here, we set the photon lifetime as $1/2\kappa=1/2\kappa_L=15$ ps, which corresponds to $Q\simeq2.0\times10^4$. The other lifetimes and the nonlinear coefficients are the same as in the main text. Namely, the carrier and thermal lifetimes are $1/\gamma_n=200$ ps and $1/\gamma_\theta=100$ ns, respectively. For the coefficients associated with optical nonlinearities, we set $f=0.0244$, $\xi=8.2\kappa/\gamma_n$, $\beta=0.0296\kappa/\gamma_\theta$, and $\eta=0.0036\kappa/\gamma_\theta$.
\begin{figure}
\includegraphics[width=0.48\textwidth]{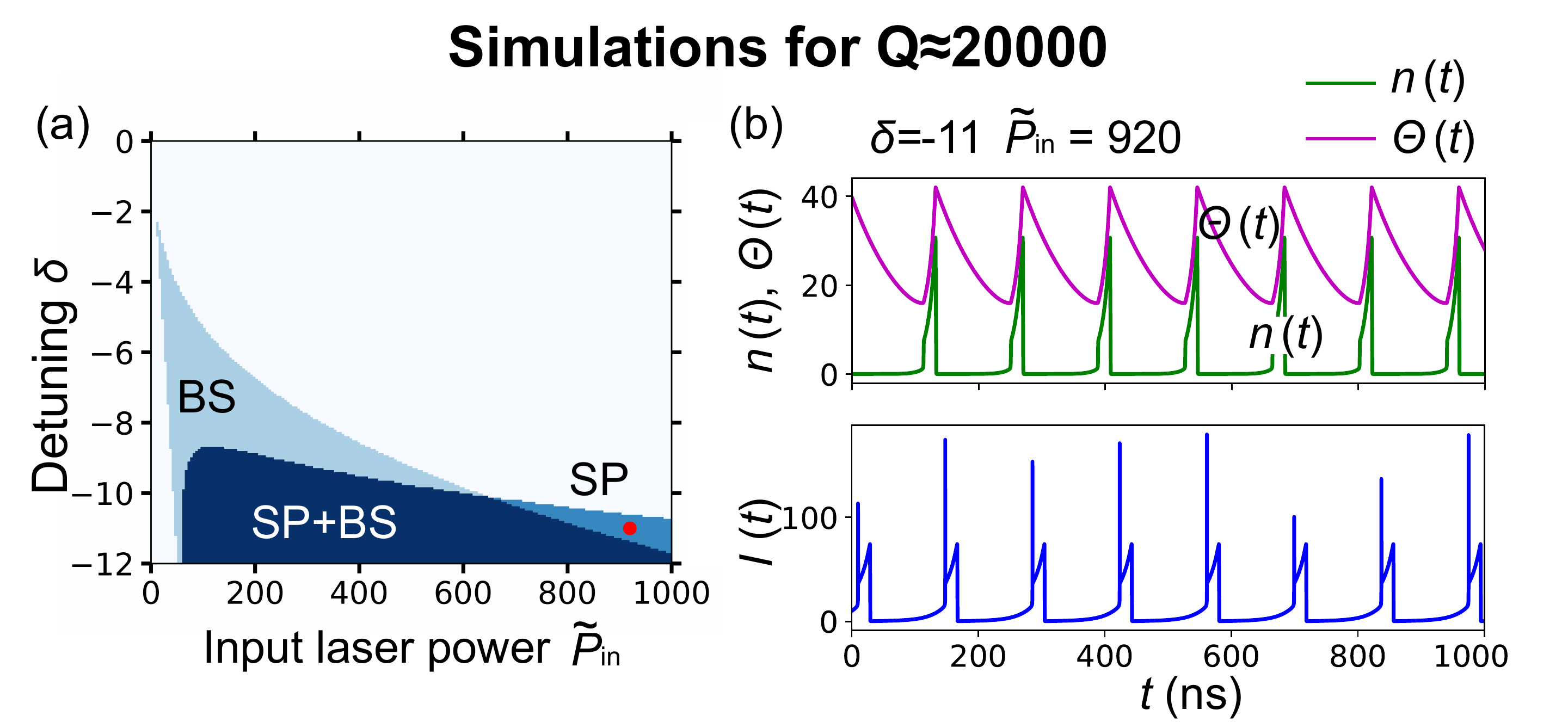}
\caption{  Simulations for a Si PhC cavity with the photon lifetime of $1/2\kappa=15$ ps with the deterministic Eqs. (\ref{eq:adiabatic_n}) and (\ref{eq:adiabatic_t}). (a) Self-pulsing (SP) and bistable (BS) regions as functions of detuning $\delta$ and scaled laser input power $\tilde{P}_{\rm in}\equiv(\kappa_L/\kappa_H)^2P_{\rm in}$, where $1/2\kappa_L$=15 ps and $1/2\kappa_H$=300 ps. (b) Real-time evolution of carrier $n(t)$, thermal effect $\theta(t)$, and light output $I(t)=|\alpha(t)|^2$ for $\delta=-11$ and $\tilde{P}_{\rm in}=920$. For simulations, we used the same parameters as in Fig. 3 in the main text except for the photon lifetime.}
\label{fig:Sup_sim}
\end{figure} 

First, we investigate the static properties as we did in the main text. In Fig. \ref{fig:Sup_sim}(a), we present the regions of bistability (BS) and self-pulsing (SP) as functions of the detuning $\delta$ and a scaled input power $\tilde{P}_{\rm in}$. Here, the detuning $\delta$ is defined with $\kappa_L$ as $\delta=(\omega_L-\omega_c)/\kappa_L$. Meanwhile, input power ${P}_{\rm in}$ is scaled as $\tilde{P}_{\rm in}\equiv(\kappa_L/\kappa_H)^2P_{\rm in}$ with $1/2\kappa_H$=300 ps for direct comparison with the input power in Fig. 3(a) in the main text. Actually, $1/2\kappa_H$=300 ps is the photon lifetime used for simulations in the main text. Figure \ref{fig:Sup_sim}(a) indicates that when the $Q$ value is decreased to around $2.0\times10^4$, the SP region shifts to a large negative detuning and largely overlaps the BS region [see the SP+BS region in Fig. \ref{fig:Sup_sim}(a)]. In the overlap (SP+BS) region, self-pulsing occurs only in the high-intensity state, which is achieved only by careful up- and down-ramping of input power without noises. Thus, in the real world, we may not be able to observe self-pulsing in the overlap (SP+BS) region. Therefore, to purely observe self-pulsing, we should use the SP region, where the detuning has a large negative value and the input power is larger than the upper threshold of bistability. To observe self-pulsing, for example, let us choose the detuning $\delta=-11$ and input power $\tilde{P}_{\rm in}=920$ (${P}_{\rm in}=2.3$), which are indicated by the filled red circle  in Fig. \ref{fig:Sup_sim}(a). This input power ($\tilde{P}_{\rm in}=$920) is a thousand times larger than the critical input power for self-pulsing in Fig. 3 in the main text and may correspond to the order of 10 mW as waveguide input power. Therefore, a higher $Q$ value dramatically reduces not only the threshold input power of bistability but also the critical input power for self-pulsing. Furthermore, since the critical point of the self-pulsing region is overlapped by bistability [see the interface between BS and SP+BS region in Fig. \ref{fig:Sup_sim}(a)], we cannot investigate the properties of limit cycle oscillation around bifurcation points.
 
Second, we simulated time evolutions of self-pulsing in the Si PhC microcavity with the moderate $Q$ value. We find that the large time scale difference $\kappa\gg\gamma_n\gg\gamma_\theta$ makes direct numerical integration of coupled-mode Eqs. (2)-(4) in the main text extremely unstable. Thus, in the same way as in Ref. \cite{VanVaerenbergh2012}, we perform adiabatic elimination of the field degree of freedom. Putting $\dot{\alpha}=0$ into Eq. (2) in the main text, we obtain
\begin{eqnarray}
\alpha=\frac{\sqrt{P_{\rm in}}}{\kappa}\cdot\frac{(1+fn)+i(-\delta-\theta+n)}{(-\delta-\theta+n)^2+(1+fn)^2}.
\label{eq:ss_alpha}
\end{eqnarray}
Substituting Eq. (\ref{eq:ss_alpha}) into Eqs. (3) and (4) in the main text, the coupled mode equations are reduced to carrier-thermal dynamics as
\begin{eqnarray}
\dot{n}&=&\gamma_n\left[-n+\left\lbrace\frac{qp}{(1+fn)^2+(-\delta-\theta+n)^2}\right\rbrace^2\right]\label{eq:adiabatic_n}\nonumber\\
\\
\dot{\theta}&=&\gamma_\theta\left[-\theta+\frac{(1+en)p}{(1+fn)^2+(-\delta-\theta+n)^2}\right],\label{eq:adiabatic_t}
\end{eqnarray} 
where we define $p\equiv\beta|\alpha_{\rm in}|^2\kappa^{-2}$, $e\equiv\eta/\beta$, and $q\equiv\sqrt{\xi}/\beta$, similarly to how it is defined in Ref. \cite{VanVaerenbergh2012}. With simulated $n(t)$ and $\theta(t)$, the electric field $\alpha(t)$ is easily calculated through Eq. (\ref{eq:ss_alpha}). We briefly comment on the validity of the adiabatic elimination in the high input power regime. In original coupled-mode Eqs. (2)-(4) in the main text, an effective field decay is given by $\kappa(1+fn)$. Thus, the adiabatic elimination condition $\kappa(1+fn)\gg\gamma_n,\gamma_\theta$ always holds if Eq. (\ref{eq:ad_condition}) is assumed.

Since the objective of this supplementary material is not a detailed study of stochastic limit cycles, we do not add noises to Eqs. (\ref{eq:adiabatic_n}) and (\ref{eq:adiabatic_t}). Figure \ref{fig:Sup_sim}(b) shows the time evolution of $n(t)$, $\theta(t)$, and $I(t)=|\alpha(t)|^2$ for $\delta=-10$ and $\tilde{P}_{\rm in}=$920 (${P}_{\rm in}=2.3$), which clearly exhibits self-pulsing (a limit cycle) with a frequency of 7.3 MHz. Interestingly, the temporal behavior of self-pulsing light output $I(t)$ shown in Fig. \ref{fig:Sup_sim}(b) is more intermittent than that shown in Fig. 3 (b) in the main text, and may resemble that reported in Ref. \cite{Brunstein2012}. Even though the temporal behavior for the moderate-Q cavity is quantitatively different from that for the high-Q cavity in the main text, as dynamical systems, both have the same bifurcation (the Hopf bifurcation) and multi-stability (bistability) structures. Namely, for both moderate- and high-Q cavities, non-trivial regions are only the three regions: the SP, BS, and SP+BS regions. 

In conclusion, simulations were performed for a Si PhC cavity with a moderate $Q\sim2.0\times10^4$. The simulations indicate that a large negative detuning and high pump power are required to observer self-pulsing in a device with a moderate $Q$. In fact, the critical input power for self-pulsing is found to be a hundred times higher than that in the measured device with a high $Q\sim3.5\times10^5$. Therefore, as we briefly explained in the main text, a high $Q$ value is technically very important for observing self-pulsing with a moderate detuning and low input power. 

\section{Carrier and thermal noises} 
Here, we consider stochastic simulations with carrier or thermal noises. In the main text, for simplicity, we added the Langevin nose to the field. It is still important to simulate the coupled-mode equations with the carrier and thermal noises.

\begin{figure}
\includegraphics[width=0.48\textwidth]{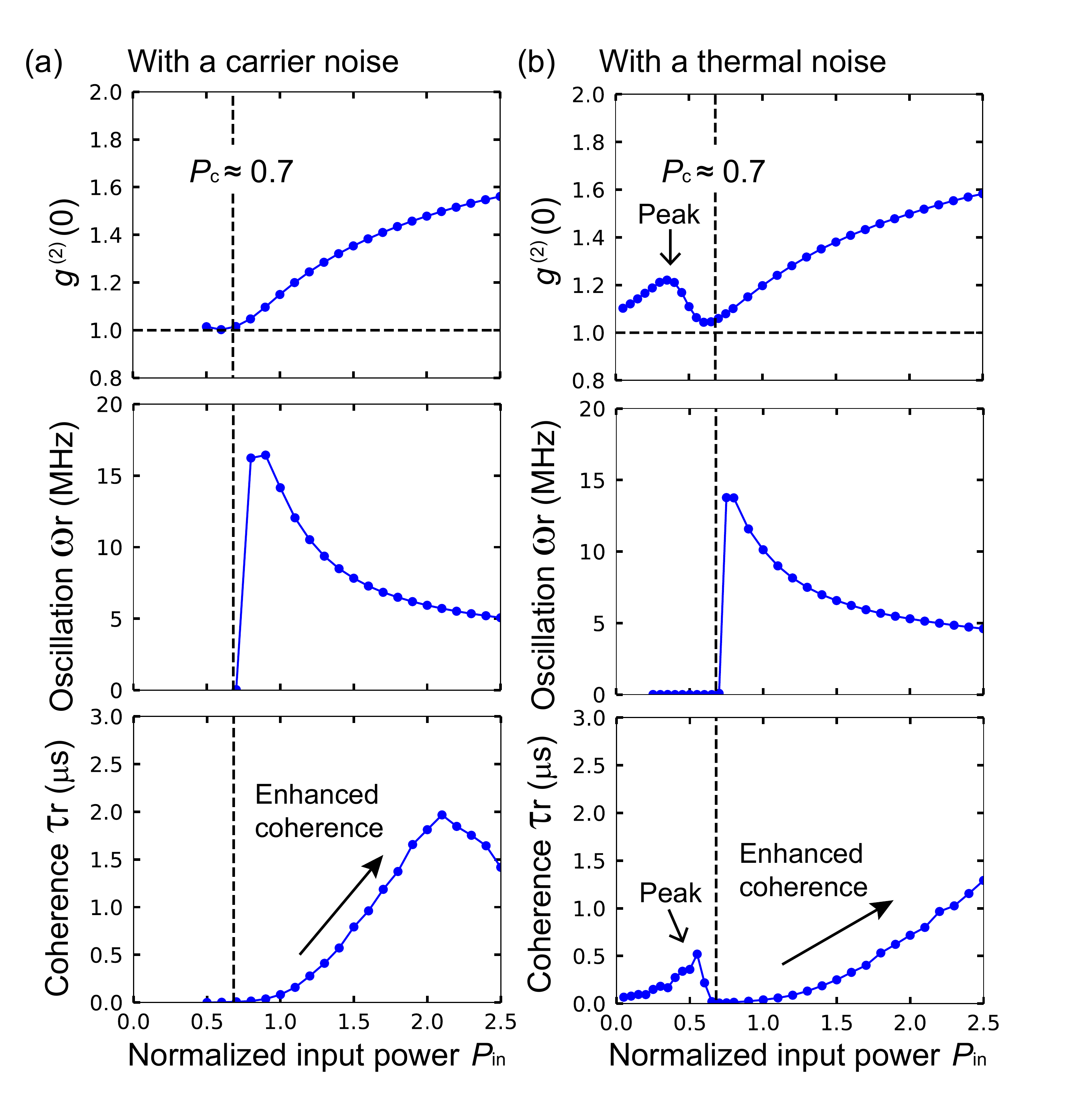}
\caption{Simulations solely with a carrier $f_n$ (a) and thermal noise $f_\theta$ (b). $g^{(2)}(0)$ (top), oscillation frequencies $\omega_{r}$ (middle), and the coherence time $\tau_r$ (bottom) of the simulated $g^{(2)}(\tau)$. All the parameters except for the noise term are the same as in Fig. 3(c) in the main text.}
\label{fig:Sup_noise}
\end{figure}

First, let us discuss an additive Langevin noise $f_n$ added to carrier dynamics [Eq. (3) in the main text]. The additive noise $f_n$ satisfies correlations
\begin{equation}
\langle f_n(t)f_n(t')\rangle=2D_n\delta_{i,j}\delta(t-t')\ \ {\rm and}\ \ \langle f_n(t)\rangle=0.
\end{equation}
The coefficient $D_n$ is the carrier noise strength. For stochastic numerical simulations, we employed the Euler-Maruyama method in the same way as in the main text. For the carrier noise strength, we set $\sqrt{2D_n}=1.1\sqrt{\kappa}$, which is much larger than the field noise strength ($\sqrt{2D_\alpha}=0.05\sqrt{\kappa}$) used in Fig. 3 in the main text. All parameters except for the noise terms are the same as in the main text. Simulation results with the carrier noise are shown in Fig. \ref{fig:Sup_noise}(a), which shows the second-order photon correlation at a zero delay time $g^{(2)}(0)$ (top), the oscillation frequency $\omega_r$ (middle), and the coherence time $\tau_r$ (bottom) of $g^{(2)}(\tau)$. We stress that the field noise is not included in the simulations in Fig. \ref{fig:Sup_noise}. Figure \ref{fig:Sup_noise}(a) is qualitatively the same as Fig. 3(c) in the main text. Namely, the oscillation frequency $\omega_r$ (middle) and coherence time $\tau_r$ (bottom) of $g^{(2)}(\tau)$ in Fig. \ref{fig:Sup_noise}(a) are almost the same as those in Fig. 3(c). Furthermore, coherence time $\tau_r$ increases after the onset of self-pulsing, and it starts to decrease in the high pump power region. Interestingly, we needed such a large value of the carrier noise strength ($\sqrt{2D_n}=1.1\sqrt{\kappa}$) to approximately reproduce the observed input power dependence of the coherence time $\tau_r$. In fact, if we use $\sqrt{2D_n}=0.05\sqrt{\kappa}$, which is the same value as the field noise strength in the main text, coherence times reach even the order of several hundreds of microseconds (not shown). These results indicates that if the field and carrier noise strengths are equal, the field noise will dominate the carrier noise. This is the reason why we did not consider the carrier noise for the simulations in the main text. However, we cannot not exclude the carrier noise as a candidate of the system's noise sources, and estimating the actual carrier noise strength is almost impossible. 

Second, we consider an additive thermal noise $f_{\theta}$ satisfying correlations
\begin{equation}
\langle f_\theta(t)f_\theta(t')\rangle=2D_\theta\delta_{i,j}\delta(t-t')\ \ {\rm and}\ \ \langle f_\theta(t)\rangle=0,
\end{equation}
where $D_\theta$ is the thermal noise strength. We use the same thermal noise strength as the field noise strength in the main text: $\sqrt{2D_\theta}=0.05\sqrt{\kappa}$. Simulations solely with the thermal noise are shown in Fig. \ref{fig:Sup_noise}(b), where the top, middle, and bottom graphs represent $g^{(2)}(0)$, the oscillation frequency $\omega_r$, and the coherence time $\tau_r$ of $g^{(2)}(\tau)$, respectively. Interestingly, both $g^{(2)}(0)$ and $\tau_r$ exhibit small peaks when input power is slightly below the critical point for self-pulsing. This peak of $g^{(2)}(0)$ is associated with the bistability of the system, and is well known in optical bistable systems \cite{Drummond1980} such as microcavity polaritons \cite{Abbaspour2015,Casteels2017,Fink2017}. Namely, the thermal noise induces jumps between the two stable states, which leads to a large intensity fluctuation and $g^{(2)}(0)>1$. The reason $g^{(2)}(0)$ did not exhibit such a peak for the simulations with the field or carrier noise may be related to the fact that our bistability is induced by the TO-induced nonlinearity. Furthermore, the peak of $\tau_r$ represents critical slowing down, which is also associated with the bistability \cite{Fink2017,Notomi2005}. Except for the peak of $g^{(2)}(0)$ and $\tau_r$ in the high input power region, all the behaviors of $g^{(2)}(0)$ and $\omega_r$ are again the same as those in Fig. 3(c) in the main text. In particular, the coherence enhancement around the onset of self-pulsing was again reproduced. Therefore, the coherence enhancement is a universal phenomenon independent of the type of noise. Since the peak of $g^{(2)}(0)$ was never been observed in our experiments, it is reasonable to conclude that the thermal noise is negligible. 

Finally, we briefly comment on the coherence drop in the high-input-power region. As we can find from Figs. \ref{fig:Sup_noise} and 3(c) in the main text, the field and carrier noises give rise to the coherence drop, while the thermal noise does not. This indicates that the origin of the coherence drop may be related to the time scale of a variable to which an additive noise is added. Namely, the coherence drop occurs in the high input power region only when an additive noise is added to a fast variable (the field or carrier variable in our model).
 
We summarize the important consequences of the argument i n this section. The first important point is that any additive noise can reproduce the coherence enhancement. Second, in order to reproduce the observed coherence time of $g^{(2)}(\tau)$, a very large carrier noise strength is required. Finally, it is reasonable to neglect the thermal noise because it cannot reproduce the coherence drop in the high-input-power region. Additionally, the thermal noise gives rise to a small peak of $g^{(2)}(0)$, which was not observed in the experiment.


\begin{thebibliography}{50}%
\makeatletter
\providecommand \@ifxundefined [1]{%
 \@ifx{#1\undefined}
}%
\providecommand \@ifnum [1]{%
 \ifnum #1\expandafter \@firstoftwo
 \else \expandafter \@secondoftwo
 \fi
}%
\providecommand \@ifx [1]{%
 \ifx #1\expandafter \@firstoftwo
 \else \expandafter \@secondoftwo
 \fi
}%
\providecommand \natexlab [1]{#1}%
\providecommand \enquote  [1]{``#1''}%
\providecommand \bibnamefont  [1]{#1}%
\providecommand \bibfnamefont [1]{#1}%
\providecommand \citenamefont [1]{#1}%
\providecommand \href@noop [0]{\@secondoftwo}%
\providecommand \href [0]{\begingroup \@sanitize@url \@href}%
\providecommand \@href[1]{\@@startlink{#1}\@@href}%
\providecommand \@@href[1]{\endgroup#1\@@endlink}%
\providecommand \@sanitize@url [0]{\catcode `\\12\catcode `\$12\catcode
  `\&12\catcode `\#12\catcode `\^12\catcode `\_12\catcode `\%12\relax}%
\providecommand \@@startlink[1]{}%
\providecommand \@@endlink[0]{}%
\providecommand \url  [0]{\begingroup\@sanitize@url \@url }%
\providecommand \@url [1]{\endgroup\@href {#1}{\urlprefix }}%
\providecommand \urlprefix  [0]{URL }%
\providecommand \Eprint [0]{\href }%
\providecommand \doibase [0]{http://dx.doi.org/}%
\providecommand \selectlanguage [0]{\@gobble}%
\providecommand \bibinfo  [0]{\@secondoftwo}%
\providecommand \bibfield  [0]{\@secondoftwo}%
\providecommand \translation [1]{[#1]}%
\providecommand \BibitemOpen [0]{}%
\providecommand \bibitemStop [0]{}%
\providecommand \bibitemNoStop [0]{.\EOS\space}%
\providecommand \EOS [0]{\spacefactor3000\relax}%
\providecommand \BibitemShut  [1]{\csname bibitem#1\endcsname}%
\let\auto@bib@innerbib\@empty
\bibitem [{\citenamefont {Nov{\'a}k}\ and\ \citenamefont
  {Tyson}(2008)}]{Novak2008}%
  \BibitemOpen
  \bibfield  {author} {\bibinfo {author} {\bibfnamefont {B.}~\bibnamefont
  {Nov{\'a}k}}\ and\ \bibinfo {author} {\bibfnamefont {J.~J.}\ \bibnamefont
  {Tyson}},\ }\href {\doibase 10.1038/nrm2530} {\bibfield  {journal} {\bibinfo
  {journal} {Nature Reviews Molecular Cell Biology}\ }\textbf {\bibinfo
  {volume} {9}},\ \bibinfo {pages} {981} (\bibinfo {year} {2008})}\BibitemShut
  {NoStop}%
\bibitem [{\citenamefont {Gaspard}(2002)}]{Gaspard2002}%
  \BibitemOpen
  \bibfield  {author} {\bibinfo {author} {\bibfnamefont {P.}~\bibnamefont
  {Gaspard}},\ }\href {\doibase 10.1063/1.1513461} {\bibfield  {journal}
  {\bibinfo  {journal} {The Journal of Chemical Physics}\ }\textbf {\bibinfo
  {volume} {117}},\ \bibinfo {pages} {8905} (\bibinfo {year} {2002})},\ \Eprint
  {http://arxiv.org/abs/https://doi.org/10.1063/1.1513461}
  {https://doi.org/10.1063/1.1513461} \BibitemShut {NoStop}%
\bibitem [{\citenamefont {Gonze}\ \emph {et~al.}(2002)\citenamefont {Gonze},
  \citenamefont {Halloy},\ and\ \citenamefont {Gaspard}}]{Gonze2002}%
  \BibitemOpen
  \bibfield  {author} {\bibinfo {author} {\bibfnamefont {D.}~\bibnamefont
  {Gonze}}, \bibinfo {author} {\bibfnamefont {J.}~\bibnamefont {Halloy}}, \
  and\ \bibinfo {author} {\bibfnamefont {P.}~\bibnamefont {Gaspard}},\ }\href
  {\doibase 10.1063/1.1475765} {\bibfield  {journal} {\bibinfo  {journal} {The
  Journal of Chemical Physics}\ }\textbf {\bibinfo {volume} {116}},\ \bibinfo
  {pages} {10997} (\bibinfo {year} {2002})},\ \Eprint
  {http://arxiv.org/abs/https://doi.org/10.1063/1.1475765}
  {https://doi.org/10.1063/1.1475765} \BibitemShut {NoStop}%
\bibitem [{\citenamefont {Qian}(2006)}]{Qian2006}%
  \BibitemOpen
  \bibfield  {author} {\bibinfo {author} {\bibfnamefont {H.}~\bibnamefont
  {Qian}},\ }\href {\doibase 10.1021/jp061858z} {\bibfield  {journal} {\bibinfo
   {journal} {The Journal of Physical Chemistry B}\ }\textbf {\bibinfo {volume}
  {110}},\ \bibinfo {pages} {15063} (\bibinfo {year} {2006})}\BibitemShut
  {NoStop}%
\bibitem [{\citenamefont {Cao}\ \emph {et~al.}(2015)\citenamefont {Cao},
  \citenamefont {Wang}, \citenamefont {Ouyang},\ and\ \citenamefont
  {Tu}}]{Cao2015}%
  \BibitemOpen
  \bibfield  {author} {\bibinfo {author} {\bibfnamefont {Y.}~\bibnamefont
  {Cao}}, \bibinfo {author} {\bibfnamefont {H.}~\bibnamefont {Wang}}, \bibinfo
  {author} {\bibfnamefont {Q.}~\bibnamefont {Ouyang}}, \ and\ \bibinfo {author}
  {\bibfnamefont {Y.}~\bibnamefont {Tu}},\ }\href
  {https://doi.org/10.1038/nphys3412} {\bibfield  {journal} {\bibinfo
  {journal} {Nature Physics}\ }\textbf {\bibinfo {volume} {11}},\ \bibinfo
  {pages} {772 EP } (\bibinfo {year} {2015})},\ \bibinfo {note}
  {article}\BibitemShut {NoStop}%
\bibitem [{\citenamefont {Barato}\ and\ \citenamefont
  {Seifert}(2016)}]{Barato2016}%
  \BibitemOpen
  \bibfield  {author} {\bibinfo {author} {\bibfnamefont {A.~C.}\ \bibnamefont
  {Barato}}\ and\ \bibinfo {author} {\bibfnamefont {U.}~\bibnamefont
  {Seifert}},\ }\href {\doibase 10.1103/PhysRevX.6.041053} {\bibfield
  {journal} {\bibinfo  {journal} {Phys. Rev. X}\ }\textbf {\bibinfo {volume}
  {6}},\ \bibinfo {pages} {041053} (\bibinfo {year} {2016})}\BibitemShut
  {NoStop}%
\bibitem [{\citenamefont {Fei}\ \emph {et~al.}(2018)\citenamefont {Fei},
  \citenamefont {Cao}, \citenamefont {Ouyang},\ and\ \citenamefont
  {Tu}}]{Fei2018}%
  \BibitemOpen
  \bibfield  {author} {\bibinfo {author} {\bibfnamefont {C.}~\bibnamefont
  {Fei}}, \bibinfo {author} {\bibfnamefont {Y.}~\bibnamefont {Cao}}, \bibinfo
  {author} {\bibfnamefont {Q.}~\bibnamefont {Ouyang}}, \ and\ \bibinfo {author}
  {\bibfnamefont {Y.}~\bibnamefont {Tu}},\ }\href {\doibase
  10.1038/s41467-018-03826-4} {\bibfield  {journal} {\bibinfo  {journal}
  {Nature Communications}\ }\textbf {\bibinfo {volume} {9}},\ \bibinfo {pages}
  {1434} (\bibinfo {year} {2018})}\BibitemShut {NoStop}%
\bibitem [{\citenamefont {Nguyen}\ \emph {et~al.}(2018)\citenamefont {Nguyen},
  \citenamefont {Seifert},\ and\ \citenamefont {Barato}}]{Nguyen2018}%
  \BibitemOpen
  \bibfield  {author} {\bibinfo {author} {\bibfnamefont {B.}~\bibnamefont
  {Nguyen}}, \bibinfo {author} {\bibfnamefont {U.}~\bibnamefont {Seifert}}, \
  and\ \bibinfo {author} {\bibfnamefont {A.~C.}\ \bibnamefont {Barato}},\
  }\href {\doibase 10.1063/1.5032104} {\bibfield  {journal} {\bibinfo
  {journal} {The Journal of Chemical Physics}\ }\textbf {\bibinfo {volume}
  {149}},\ \bibinfo {pages} {045101} (\bibinfo {year} {2018})},\ \Eprint
  {http://arxiv.org/abs/https://doi.org/10.1063/1.5032104}
  {https://doi.org/10.1063/1.5032104} \BibitemShut {NoStop}%
\bibitem [{\citenamefont {Kuramoto}(2003)}]{Kuramoto2003}%
  \BibitemOpen
  \bibfield  {author} {\bibinfo {author} {\bibfnamefont {Y.}~\bibnamefont
  {Kuramoto}},\ }\href@noop {} {\emph {\bibinfo {title} {Chemical oscillations,
  waves, and turbulence}}}\ (\bibinfo  {publisher} {Courier Corporation},\
  \bibinfo {year} {2003})\BibitemShut {NoStop}%
\bibitem [{\citenamefont {Nakao}(2017)}]{Nakao2017}%
  \BibitemOpen
  \bibfield  {author} {\bibinfo {author} {\bibfnamefont {H.}~\bibnamefont
  {Nakao}},\ }\href {\doibase 10.1080/00107514.2015.1094987} {\  (\bibinfo
  {year} {2017}),\ 10.1080/00107514.2015.1094987},\ \Eprint
  {http://arxiv.org/abs/arXiv:1704.03293} {arXiv:1704.03293} \BibitemShut
  {NoStop}%
\bibitem [{\citenamefont {Barclay}\ \emph {et~al.}(2005)\citenamefont
  {Barclay}, \citenamefont {Srinivasan},\ and\ \citenamefont
  {Painter}}]{Barclay2005}%
  \BibitemOpen
  \bibfield  {author} {\bibinfo {author} {\bibfnamefont {P.~E.}\ \bibnamefont
  {Barclay}}, \bibinfo {author} {\bibfnamefont {K.}~\bibnamefont {Srinivasan}},
  \ and\ \bibinfo {author} {\bibfnamefont {O.}~\bibnamefont {Painter}},\ }\href
  {\doibase 10.1364/OPEX.13.000801} {\bibfield  {journal} {\bibinfo  {journal}
  {Opt. Express}\ }\textbf {\bibinfo {volume} {13}},\ \bibinfo {pages} {801}
  (\bibinfo {year} {2005})}\BibitemShut {NoStop}%
\bibitem [{\citenamefont {Uesugi}\ \emph {et~al.}(2006)\citenamefont {Uesugi},
  \citenamefont {Song}, \citenamefont {Asano},\ and\ \citenamefont
  {Noda}}]{Uesugi2006}%
  \BibitemOpen
  \bibfield  {author} {\bibinfo {author} {\bibfnamefont {T.}~\bibnamefont
  {Uesugi}}, \bibinfo {author} {\bibfnamefont {B.-S.}\ \bibnamefont {Song}},
  \bibinfo {author} {\bibfnamefont {T.}~\bibnamefont {Asano}}, \ and\ \bibinfo
  {author} {\bibfnamefont {S.}~\bibnamefont {Noda}},\ }\href {\doibase
  10.1364/OPEX.14.000377} {\bibfield  {journal} {\bibinfo  {journal} {Opt.
  Express}\ }\textbf {\bibinfo {volume} {14}},\ \bibinfo {pages} {377}
  (\bibinfo {year} {2006})}\BibitemShut {NoStop}%
\bibitem [{\citenamefont {Leuthold}\ \emph {et~al.}(2010)\citenamefont
  {Leuthold}, \citenamefont {Koos},\ and\ \citenamefont
  {Freude}}]{Leuthold2010}%
  \BibitemOpen
  \bibfield  {author} {\bibinfo {author} {\bibfnamefont {J.}~\bibnamefont
  {Leuthold}}, \bibinfo {author} {\bibfnamefont {C.}~\bibnamefont {Koos}}, \
  and\ \bibinfo {author} {\bibfnamefont {W.}~\bibnamefont {Freude}},\ }\href
  {https://doi.org/10.1038/nphoton.2010.185} {\bibfield  {journal} {\bibinfo
  {journal} {Nature Photonics}\ }\textbf {\bibinfo {volume} {4}},\ \bibinfo
  {pages} {535 EP } (\bibinfo {year} {2010})},\ \bibinfo {note} {review
  Article}\BibitemShut {NoStop}%
\bibitem [{\citenamefont {Tanabe}\ \emph {et~al.}(2005)\citenamefont {Tanabe},
  \citenamefont {Notomi}, \citenamefont {Mitsugi}, \citenamefont {Shinya},\
  and\ \citenamefont {Kuramochi}}]{Tanabe2005}%
  \BibitemOpen
  \bibfield  {author} {\bibinfo {author} {\bibfnamefont {T.}~\bibnamefont
  {Tanabe}}, \bibinfo {author} {\bibfnamefont {M.}~\bibnamefont {Notomi}},
  \bibinfo {author} {\bibfnamefont {S.}~\bibnamefont {Mitsugi}}, \bibinfo
  {author} {\bibfnamefont {A.}~\bibnamefont {Shinya}}, \ and\ \bibinfo {author}
  {\bibfnamefont {E.}~\bibnamefont {Kuramochi}},\ }\href {\doibase
  10.1364/OL.30.002575} {\bibfield  {journal} {\bibinfo  {journal} {Opt.
  Lett.}\ }\textbf {\bibinfo {volume} {30}},\ \bibinfo {pages} {2575} (\bibinfo
  {year} {2005})}\BibitemShut {NoStop}%
\bibitem [{\citenamefont {Notomi}\ \emph {et~al.}(2005)\citenamefont {Notomi},
  \citenamefont {Shinya}, \citenamefont {Mitsugi}, \citenamefont {Kira},
  \citenamefont {Kuramochi},\ and\ \citenamefont {Tanabe}}]{Notomi2005}%
  \BibitemOpen
  \bibfield  {author} {\bibinfo {author} {\bibfnamefont {M.}~\bibnamefont
  {Notomi}}, \bibinfo {author} {\bibfnamefont {A.}~\bibnamefont {Shinya}},
  \bibinfo {author} {\bibfnamefont {S.}~\bibnamefont {Mitsugi}}, \bibinfo
  {author} {\bibfnamefont {G.}~\bibnamefont {Kira}}, \bibinfo {author}
  {\bibfnamefont {E.}~\bibnamefont {Kuramochi}}, \ and\ \bibinfo {author}
  {\bibfnamefont {T.}~\bibnamefont {Tanabe}},\ }\href {\doibase
  10.1364/OPEX.13.002678} {\bibfield  {journal} {\bibinfo  {journal} {Opt.
  Express}\ }\textbf {\bibinfo {volume} {13}},\ \bibinfo {pages} {2678}
  (\bibinfo {year} {2005})}\BibitemShut {NoStop}%
\bibitem [{\citenamefont {Tanabe}\ \emph
  {et~al.}(2007{\natexlab{a}})\citenamefont {Tanabe}, \citenamefont {Shinya},
  \citenamefont {Kuramochi}, \citenamefont {Kondo}, \citenamefont {Taniyama},\
  and\ \citenamefont {Notomi}}]{Tanabe2007}%
  \BibitemOpen
  \bibfield  {author} {\bibinfo {author} {\bibfnamefont {T.}~\bibnamefont
  {Tanabe}}, \bibinfo {author} {\bibfnamefont {A.}~\bibnamefont {Shinya}},
  \bibinfo {author} {\bibfnamefont {E.}~\bibnamefont {Kuramochi}}, \bibinfo
  {author} {\bibfnamefont {S.}~\bibnamefont {Kondo}}, \bibinfo {author}
  {\bibfnamefont {H.}~\bibnamefont {Taniyama}}, \ and\ \bibinfo {author}
  {\bibfnamefont {M.}~\bibnamefont {Notomi}},\ }\href {\doibase
  10.1063/1.2757099} {\bibfield  {journal} {\bibinfo  {journal} {Applied
  Physics Letters}\ }\textbf {\bibinfo {volume} {91}},\ \bibinfo {pages}
  {021110} (\bibinfo {year} {2007}{\natexlab{a}})},\ \Eprint
  {http://arxiv.org/abs/https://doi.org/10.1063/1.2757099}
  {https://doi.org/10.1063/1.2757099} \BibitemShut {NoStop}%
\bibitem [{\citenamefont {Weidner}\ \emph {et~al.}(2007)\citenamefont
  {Weidner}, \citenamefont {Combrié}, \citenamefont {de~Rossi}, \citenamefont
  {Tran},\ and\ \citenamefont {Cassette}}]{Weidner2007}%
  \BibitemOpen
  \bibfield  {author} {\bibinfo {author} {\bibfnamefont {E.}~\bibnamefont
  {Weidner}}, \bibinfo {author} {\bibfnamefont {S.}~\bibnamefont {Combrié}},
  \bibinfo {author} {\bibfnamefont {A.}~\bibnamefont {de~Rossi}}, \bibinfo
  {author} {\bibfnamefont {N.-V.-Q.}\ \bibnamefont {Tran}}, \ and\ \bibinfo
  {author} {\bibfnamefont {S.}~\bibnamefont {Cassette}},\ }\href {\doibase
  10.1063/1.2712502} {\bibfield  {journal} {\bibinfo  {journal} {Applied
  Physics Letters}\ }\textbf {\bibinfo {volume} {90}},\ \bibinfo {pages}
  {101118} (\bibinfo {year} {2007})},\ \Eprint
  {http://arxiv.org/abs/https://doi.org/10.1063/1.2712502}
  {https://doi.org/10.1063/1.2712502} \BibitemShut {NoStop}%
\bibitem [{\citenamefont {Haret}\ \emph {et~al.}(2009)\citenamefont {Haret},
  \citenamefont {Tanabe}, \citenamefont {Kuramochi},\ and\ \citenamefont
  {Notomi}}]{Haret2009}%
  \BibitemOpen
  \bibfield  {author} {\bibinfo {author} {\bibfnamefont {L.-D.}\ \bibnamefont
  {Haret}}, \bibinfo {author} {\bibfnamefont {T.}~\bibnamefont {Tanabe}},
  \bibinfo {author} {\bibfnamefont {E.}~\bibnamefont {Kuramochi}}, \ and\
  \bibinfo {author} {\bibfnamefont {M.}~\bibnamefont {Notomi}},\ }\href
  {\doibase 10.1364/OE.17.021108} {\bibfield  {journal} {\bibinfo  {journal}
  {Opt. Express}\ }\textbf {\bibinfo {volume} {17}},\ \bibinfo {pages} {21108}
  (\bibinfo {year} {2009})}\BibitemShut {NoStop}%
\bibitem [{\citenamefont {de~Rossi}\ \emph {et~al.}(2009)\citenamefont
  {de~Rossi}, \citenamefont {Lauritano}, \citenamefont {Combri\'e},
  \citenamefont {Tran},\ and\ \citenamefont {Husko}}]{Rossi2009}%
  \BibitemOpen
  \bibfield  {author} {\bibinfo {author} {\bibfnamefont {A.}~\bibnamefont
  {de~Rossi}}, \bibinfo {author} {\bibfnamefont {M.}~\bibnamefont {Lauritano}},
  \bibinfo {author} {\bibfnamefont {S.}~\bibnamefont {Combri\'e}}, \bibinfo
  {author} {\bibfnamefont {Q.~V.}\ \bibnamefont {Tran}}, \ and\ \bibinfo
  {author} {\bibfnamefont {C.}~\bibnamefont {Husko}},\ }\href {\doibase
  10.1103/PhysRevA.79.043818} {\bibfield  {journal} {\bibinfo  {journal} {Phys.
  Rev. A}\ }\textbf {\bibinfo {volume} {79}},\ \bibinfo {pages} {043818}
  (\bibinfo {year} {2009})}\BibitemShut {NoStop}%
\bibitem [{\citenamefont {Priem}\ \emph {et~al.}(2005)\citenamefont {Priem},
  \citenamefont {Dumon}, \citenamefont {Bogaerts}, \citenamefont {Thourhout},
  \citenamefont {Morthier},\ and\ \citenamefont {Baets}}]{Priem2005}%
  \BibitemOpen
  \bibfield  {author} {\bibinfo {author} {\bibfnamefont {G.}~\bibnamefont
  {Priem}}, \bibinfo {author} {\bibfnamefont {P.}~\bibnamefont {Dumon}},
  \bibinfo {author} {\bibfnamefont {W.}~\bibnamefont {Bogaerts}}, \bibinfo
  {author} {\bibfnamefont {D.~V.}\ \bibnamefont {Thourhout}}, \bibinfo {author}
  {\bibfnamefont {G.}~\bibnamefont {Morthier}}, \ and\ \bibinfo {author}
  {\bibfnamefont {R.}~\bibnamefont {Baets}},\ }\href {\doibase
  10.1364/OPEX.13.009623} {\bibfield  {journal} {\bibinfo  {journal} {Opt.
  Express}\ }\textbf {\bibinfo {volume} {13}},\ \bibinfo {pages} {9623}
  (\bibinfo {year} {2005})}\BibitemShut {NoStop}%
\bibitem [{\citenamefont {Johnson}\ \emph {et~al.}(2006)\citenamefont
  {Johnson}, \citenamefont {Borselli},\ and\ \citenamefont
  {Painter}}]{Johnson2006}%
  \BibitemOpen
  \bibfield  {author} {\bibinfo {author} {\bibfnamefont {T.~J.}\ \bibnamefont
  {Johnson}}, \bibinfo {author} {\bibfnamefont {M.}~\bibnamefont {Borselli}}, \
  and\ \bibinfo {author} {\bibfnamefont {O.}~\bibnamefont {Painter}},\ }\href
  {\doibase 10.1364/OPEX.14.000817} {\bibfield  {journal} {\bibinfo  {journal}
  {Opt. Express}\ }\textbf {\bibinfo {volume} {14}},\ \bibinfo {pages} {817}
  (\bibinfo {year} {2006})}\BibitemShut {NoStop}%
\bibitem [{\citenamefont {Pernice}\ \emph {et~al.}(2010)\citenamefont
  {Pernice}, \citenamefont {Li},\ and\ \citenamefont {Tang}}]{Pernice2010}%
  \BibitemOpen
  \bibfield  {author} {\bibinfo {author} {\bibfnamefont {W.~H.~P.}\
  \bibnamefont {Pernice}}, \bibinfo {author} {\bibfnamefont {M.}~\bibnamefont
  {Li}}, \ and\ \bibinfo {author} {\bibfnamefont {H.~X.}\ \bibnamefont
  {Tang}},\ }\href {\doibase 10.1364/OE.18.018438} {\bibfield  {journal}
  {\bibinfo  {journal} {Opt. Express}\ }\textbf {\bibinfo {volume} {18}},\
  \bibinfo {pages} {18438} (\bibinfo {year} {2010})}\BibitemShut {NoStop}%
\bibitem [{\citenamefont {Malaguti}\ \emph {et~al.}(2011)\citenamefont
  {Malaguti}, \citenamefont {Bellanca}, \citenamefont {de~Rossi}, \citenamefont
  {Combri\'e},\ and\ \citenamefont {Trillo}}]{Malaguti2011}%
  \BibitemOpen
  \bibfield  {author} {\bibinfo {author} {\bibfnamefont {S.}~\bibnamefont
  {Malaguti}}, \bibinfo {author} {\bibfnamefont {G.}~\bibnamefont {Bellanca}},
  \bibinfo {author} {\bibfnamefont {A.}~\bibnamefont {de~Rossi}}, \bibinfo
  {author} {\bibfnamefont {S.}~\bibnamefont {Combri\'e}}, \ and\ \bibinfo
  {author} {\bibfnamefont {S.}~\bibnamefont {Trillo}},\ }\href {\doibase
  10.1103/PhysRevA.83.051802} {\bibfield  {journal} {\bibinfo  {journal} {Phys.
  Rev. A}\ }\textbf {\bibinfo {volume} {83}},\ \bibinfo {pages} {051802}
  (\bibinfo {year} {2011})}\BibitemShut {NoStop}%
\bibitem [{\citenamefont {Cazier}\ \emph {et~al.}(2013)\citenamefont {Cazier},
  \citenamefont {Checoury}, \citenamefont {Haret},\ and\ \citenamefont
  {Boucaud}}]{Cazier2013}%
  \BibitemOpen
  \bibfield  {author} {\bibinfo {author} {\bibfnamefont {N.}~\bibnamefont
  {Cazier}}, \bibinfo {author} {\bibfnamefont {X.}~\bibnamefont {Checoury}},
  \bibinfo {author} {\bibfnamefont {L.-D.}\ \bibnamefont {Haret}}, \ and\
  \bibinfo {author} {\bibfnamefont {P.}~\bibnamefont {Boucaud}},\ }\href
  {\doibase 10.1364/OE.21.013626} {\bibfield  {journal} {\bibinfo  {journal}
  {Opt. Express}\ }\textbf {\bibinfo {volume} {21}},\ \bibinfo {pages} {13626}
  (\bibinfo {year} {2013})}\BibitemShut {NoStop}%
\bibitem [{\citenamefont {Yacomotti}\ \emph {et~al.}(2013)\citenamefont
  {Yacomotti}, \citenamefont {Haddadi},\ and\ \citenamefont
  {Barbay}}]{Yacomotti2013}%
  \BibitemOpen
  \bibfield  {author} {\bibinfo {author} {\bibfnamefont {A.~M.}\ \bibnamefont
  {Yacomotti}}, \bibinfo {author} {\bibfnamefont {S.}~\bibnamefont {Haddadi}},
  \ and\ \bibinfo {author} {\bibfnamefont {S.}~\bibnamefont {Barbay}},\ }\href
  {\doibase 10.1103/PhysRevA.87.041804} {\bibfield  {journal} {\bibinfo
  {journal} {Phys. Rev. A}\ }\textbf {\bibinfo {volume} {87}},\ \bibinfo
  {pages} {041804} (\bibinfo {year} {2013})}\BibitemShut {NoStop}%
\bibitem [{\citenamefont {Yu}\ \emph {et~al.}(2017)\citenamefont {Yu},
  \citenamefont {Xue}, \citenamefont {Semenova}, \citenamefont {Yvind},\ and\
  \citenamefont {Mork}}]{Yu2017}%
  \BibitemOpen
  \bibfield  {author} {\bibinfo {author} {\bibfnamefont {Y.}~\bibnamefont
  {Yu}}, \bibinfo {author} {\bibfnamefont {W.}~\bibnamefont {Xue}}, \bibinfo
  {author} {\bibfnamefont {E.}~\bibnamefont {Semenova}}, \bibinfo {author}
  {\bibfnamefont {K.}~\bibnamefont {Yvind}}, \ and\ \bibinfo {author}
  {\bibfnamefont {J.}~\bibnamefont {Mork}},\ }\href {\doibase
  10.1038/nphoton.2016.248} {\bibfield  {journal} {\bibinfo  {journal} {Nature
  Photonics}\ }\textbf {\bibinfo {volume} {11}},\ \bibinfo {pages} {81}
  (\bibinfo {year} {2017})}\BibitemShut {NoStop}%
\bibitem [{\citenamefont {Yacomotti}\ \emph {et~al.}(2006)\citenamefont
  {Yacomotti}, \citenamefont {Monnier}, \citenamefont {Raineri}, \citenamefont
  {Bakir}, \citenamefont {Seassal}, \citenamefont {Raj},\ and\ \citenamefont
  {Levenson}}]{Yacomotti2006}%
  \BibitemOpen
  \bibfield  {author} {\bibinfo {author} {\bibfnamefont {A.~M.}\ \bibnamefont
  {Yacomotti}}, \bibinfo {author} {\bibfnamefont {P.}~\bibnamefont {Monnier}},
  \bibinfo {author} {\bibfnamefont {F.}~\bibnamefont {Raineri}}, \bibinfo
  {author} {\bibfnamefont {B.~B.}\ \bibnamefont {Bakir}}, \bibinfo {author}
  {\bibfnamefont {C.}~\bibnamefont {Seassal}}, \bibinfo {author} {\bibfnamefont
  {R.}~\bibnamefont {Raj}}, \ and\ \bibinfo {author} {\bibfnamefont {J.~A.}\
  \bibnamefont {Levenson}},\ }\href {\doibase 10.1103/PhysRevLett.97.143904}
  {\bibfield  {journal} {\bibinfo  {journal} {Phys. Rev. Lett.}\ }\textbf
  {\bibinfo {volume} {97}},\ \bibinfo {pages} {143904} (\bibinfo {year}
  {2006})}\BibitemShut {NoStop}%
\bibitem [{\citenamefont {Brunstein}\ \emph {et~al.}(2012)\citenamefont
  {Brunstein}, \citenamefont {Yacomotti}, \citenamefont {Sagnes}, \citenamefont
  {Raineri}, \citenamefont {Bigot},\ and\ \citenamefont
  {Levenson}}]{Brunstein2012}%
  \BibitemOpen
  \bibfield  {author} {\bibinfo {author} {\bibfnamefont {M.}~\bibnamefont
  {Brunstein}}, \bibinfo {author} {\bibfnamefont {A.~M.}\ \bibnamefont
  {Yacomotti}}, \bibinfo {author} {\bibfnamefont {I.}~\bibnamefont {Sagnes}},
  \bibinfo {author} {\bibfnamefont {F.}~\bibnamefont {Raineri}}, \bibinfo
  {author} {\bibfnamefont {L.}~\bibnamefont {Bigot}}, \ and\ \bibinfo {author}
  {\bibfnamefont {A.}~\bibnamefont {Levenson}},\ }\href {\doibase
  10.1103/PhysRevA.85.031803} {\bibfield  {journal} {\bibinfo  {journal} {Phys.
  Rev. A}\ }\textbf {\bibinfo {volume} {85}},\ \bibinfo {pages} {031803}
  (\bibinfo {year} {2012})}\BibitemShut {NoStop}%
\bibitem [{\citenamefont {Kuramochi}\ \emph {et~al.}(2014)\citenamefont
  {Kuramochi}, \citenamefont {Grossman}, \citenamefont {Nozaki}, \citenamefont
  {Takeda}, \citenamefont {Shinya}, \citenamefont {Taniyama},\ and\
  \citenamefont {Notomi}}]{Kuramochi2014}%
  \BibitemOpen
  \bibfield  {author} {\bibinfo {author} {\bibfnamefont {E.}~\bibnamefont
  {Kuramochi}}, \bibinfo {author} {\bibfnamefont {E.}~\bibnamefont {Grossman}},
  \bibinfo {author} {\bibfnamefont {K.}~\bibnamefont {Nozaki}}, \bibinfo
  {author} {\bibfnamefont {K.}~\bibnamefont {Takeda}}, \bibinfo {author}
  {\bibfnamefont {A.}~\bibnamefont {Shinya}}, \bibinfo {author} {\bibfnamefont
  {H.}~\bibnamefont {Taniyama}}, \ and\ \bibinfo {author} {\bibfnamefont
  {M.}~\bibnamefont {Notomi}},\ }\href {\doibase 10.1364/OL.39.005780}
  {\bibfield  {journal} {\bibinfo  {journal} {Opt. Lett.}\ }\textbf {\bibinfo
  {volume} {39}},\ \bibinfo {pages} {5780} (\bibinfo {year}
  {2014})}\BibitemShut {NoStop}%
\bibitem [{\citenamefont {Tanabe}\ \emph
  {et~al.}(2007{\natexlab{b}})\citenamefont {Tanabe}, \citenamefont {Notomi},
  \citenamefont {Kuramochi}, \citenamefont {Shinya},\ and\ \citenamefont
  {Taniyama}}]{Tanabe2007a}%
  \BibitemOpen
  \bibfield  {author} {\bibinfo {author} {\bibfnamefont {T.}~\bibnamefont
  {Tanabe}}, \bibinfo {author} {\bibfnamefont {M.}~\bibnamefont {Notomi}},
  \bibinfo {author} {\bibfnamefont {E.}~\bibnamefont {Kuramochi}}, \bibinfo
  {author} {\bibfnamefont {A.}~\bibnamefont {Shinya}}, \ and\ \bibinfo {author}
  {\bibfnamefont {H.}~\bibnamefont {Taniyama}},\ }\href {\doibase
  10.1038/nphoton.2006.51} {\bibfield  {journal} {\bibinfo  {journal} {Nature
  Photonics}\ }\textbf {\bibinfo {volume} {1}},\ \bibinfo {pages} {49}
  (\bibinfo {year} {2007}{\natexlab{b}})}\BibitemShut {NoStop}%
\bibitem [{\citenamefont {Mandel}\ and\ \citenamefont
  {Wolf}(1995)}]{Mandel1995}%
  \BibitemOpen
  \bibfield  {author} {\bibinfo {author} {\bibfnamefont {L.}~\bibnamefont
  {Mandel}}\ and\ \bibinfo {author} {\bibfnamefont {E.}~\bibnamefont {Wolf}},\
  }\href@noop {} {\emph {\bibinfo {title} {Optical coherence and quantum
  optics}}}\ (\bibinfo  {publisher} {Cambridge university press},\ \bibinfo
  {year} {1995})\BibitemShut {NoStop}%
\bibitem [{\citenamefont {Loudon}(1980)}]{Loudon1980}%
  \BibitemOpen
  \bibfield  {author} {\bibinfo {author} {\bibfnamefont {R.}~\bibnamefont
  {Loudon}},\ }\href {\doibase 10.1088/0034-4885/43/7/002} {\bibfield
  {journal} {\bibinfo  {journal} {Reports on Progress in Physics}\ }\textbf
  {\bibinfo {volume} {43}},\ \bibinfo {pages} {913} (\bibinfo {year}
  {1980})}\BibitemShut {NoStop}%
\bibitem [{\citenamefont {Van~Vaerenbergh}\ \emph {et~al.}(2012)\citenamefont
  {Van~Vaerenbergh}, \citenamefont {Fiers}, \citenamefont {Dambre},\ and\
  \citenamefont {Bienstman}}]{VanVaerenbergh2012}%
  \BibitemOpen
  \bibfield  {author} {\bibinfo {author} {\bibfnamefont {T.}~\bibnamefont
  {Van~Vaerenbergh}}, \bibinfo {author} {\bibfnamefont {M.}~\bibnamefont
  {Fiers}}, \bibinfo {author} {\bibfnamefont {J.}~\bibnamefont {Dambre}}, \
  and\ \bibinfo {author} {\bibfnamefont {P.}~\bibnamefont {Bienstman}},\ }\href
  {\doibase 10.1103/PhysRevA.86.063808} {\bibfield  {journal} {\bibinfo
  {journal} {Phys. Rev. A}\ }\textbf {\bibinfo {volume} {86}},\ \bibinfo
  {pages} {063808} (\bibinfo {year} {2012})}\BibitemShut {NoStop}%
\bibitem [{\citenamefont {Zhang}\ \emph {et~al.}(2013)\citenamefont {Zhang},
  \citenamefont {Fei}, \citenamefont {Cao}, \citenamefont {Cao}, \citenamefont
  {Xu},\ and\ \citenamefont {Chen}}]{Zhang2013}%
  \BibitemOpen
  \bibfield  {author} {\bibinfo {author} {\bibfnamefont {L.}~\bibnamefont
  {Zhang}}, \bibinfo {author} {\bibfnamefont {Y.}~\bibnamefont {Fei}}, \bibinfo
  {author} {\bibfnamefont {T.}~\bibnamefont {Cao}}, \bibinfo {author}
  {\bibfnamefont {Y.}~\bibnamefont {Cao}}, \bibinfo {author} {\bibfnamefont
  {Q.}~\bibnamefont {Xu}}, \ and\ \bibinfo {author} {\bibfnamefont
  {S.}~\bibnamefont {Chen}},\ }\href {\doibase 10.1103/PhysRevA.87.053805}
  {\bibfield  {journal} {\bibinfo  {journal} {Phys. Rev. A}\ }\textbf {\bibinfo
  {volume} {87}},\ \bibinfo {pages} {053805} (\bibinfo {year}
  {2013})}\BibitemShut {NoStop}%
\bibitem [{\citenamefont {Tanabe}\ \emph {et~al.}(2008)\citenamefont {Tanabe},
  \citenamefont {Taniyama},\ and\ \citenamefont {Notomi}}]{Tanabe2008}%
  \BibitemOpen
  \bibfield  {author} {\bibinfo {author} {\bibfnamefont {T.}~\bibnamefont
  {Tanabe}}, \bibinfo {author} {\bibfnamefont {H.}~\bibnamefont {Taniyama}}, \
  and\ \bibinfo {author} {\bibfnamefont {M.}~\bibnamefont {Notomi}},\ }\href
  {http://jlt.osa.org/abstract.cfm?URI=jlt-26-11-1396} {\bibfield  {journal}
  {\bibinfo  {journal} {J. Lightwave Technol.}\ }\textbf {\bibinfo {volume}
  {26}},\ \bibinfo {pages} {1396} (\bibinfo {year} {2008})}\BibitemShut
  {NoStop}%
\bibitem [{\citenamefont {Strogatz}(2018)}]{Strogatz2018}%
  \BibitemOpen
  \bibfield  {author} {\bibinfo {author} {\bibfnamefont {S.~H.}\ \bibnamefont
  {Strogatz}},\ }\href@noop {} {\emph {\bibinfo {title} {Nonlinear dynamics and
  chaos: with applications to physics, biology, chemistry, and engineering}}}\
  (\bibinfo  {publisher} {CRC Press},\ \bibinfo {year} {2018})\BibitemShut
  {NoStop}%
\bibitem [{Note1()}]{Note1}%
  \BibitemOpen
  \bibinfo {note} {In actual numerical simulations, we introduce the noise to a
  difference equation as $\sigma \xi _R\protect \sqrt {dt}$, where $\sigma
  =0.05\protect \sqrt {\kappa }$ and $\xi _R$ is a random umber following the
  normal distribution $N(0,1)$.}\BibitemShut {Stop}%
\bibitem [{\citenamefont {Louisell}(1973)}]{Louisell1973}%
  \BibitemOpen
  \bibfield  {author} {\bibinfo {author} {\bibnamefont {Louisell}},\
  }\href@noop {} {\emph {\bibinfo {title} {Quantum statistical properties of
  radiation}}},\ Vol.~\bibinfo {volume} {7}\ (\bibinfo  {publisher} {Wiley New
  York},\ \bibinfo {year} {1973})\BibitemShut {NoStop}%
\bibitem [{\citenamefont {Van~Kampen}(1992)}]{VanKampen1992}%
  \BibitemOpen
  \bibfield  {author} {\bibinfo {author} {\bibfnamefont {N.~G.}\ \bibnamefont
  {Van~Kampen}},\ }\href@noop {} {\emph {\bibinfo {title} {Stochastic processes
  in physics and chemistry}}},\ Vol.~\bibinfo {volume} {1}\ (\bibinfo
  {publisher} {Elsevier},\ \bibinfo {year} {1992})\BibitemShut {NoStop}%
\bibitem [{\citenamefont {Risken}(1996)}]{Risken1996}%
  \BibitemOpen
  \bibfield  {author} {\bibinfo {author} {\bibfnamefont {H.}~\bibnamefont
  {Risken}},\ }in\ \href@noop {} {\emph {\bibinfo {booktitle} {The
  Fokker-Planck Equation}}}\ (\bibinfo  {publisher} {Springer},\ \bibinfo
  {year} {1996})\ pp.\ \bibinfo {pages} {63--95}\BibitemShut {NoStop}%
\bibitem [{Note2()}]{Note2}%
  \BibitemOpen
  \bibinfo {note} {One may find that Eq. (\ref {eq:diffusion}) is analogous to
  the well-known Schawlow-Townes linewidth reduction in laser
  physics.}\BibitemShut {Stop}%
\bibitem [{\citenamefont {Scully}\ and\ \citenamefont
  {Zubairy}(1999)}]{Scully1999}%
  \BibitemOpen
  \bibfield  {author} {\bibinfo {author} {\bibfnamefont {M.~O.}\ \bibnamefont
  {Scully}}\ and\ \bibinfo {author} {\bibfnamefont {M.~S.}\ \bibnamefont
  {Zubairy}},\ }\href@noop {} {\enquote {\bibinfo {title} {Quantum optics},}\ }
  (\bibinfo {year} {1999})\BibitemShut {NoStop}%
\bibitem [{\citenamefont {Marconi}\ \emph {et~al.}(2019)\citenamefont
  {Marconi}, \citenamefont {Raineri}, \citenamefont {Levenson}, \citenamefont
  {Yacomotti}, \citenamefont {Javaloyes}, \citenamefont {Pan}, \citenamefont
  {Amili},\ and\ \citenamefont {Fainman}}]{Marconi2019}%
  \BibitemOpen
  \bibfield  {author} {\bibinfo {author} {\bibfnamefont {M.}~\bibnamefont
  {Marconi}}, \bibinfo {author} {\bibfnamefont {F.}~\bibnamefont {Raineri}},
  \bibinfo {author} {\bibfnamefont {A.}~\bibnamefont {Levenson}}, \bibinfo
  {author} {\bibfnamefont {A.~M.}\ \bibnamefont {Yacomotti}}, \bibinfo {author}
  {\bibfnamefont {J.}~\bibnamefont {Javaloyes}}, \bibinfo {author}
  {\bibfnamefont {S.~H.}\ \bibnamefont {Pan}}, \bibinfo {author} {\bibfnamefont
  {A.~E.}\ \bibnamefont {Amili}}, \ and\ \bibinfo {author} {\bibfnamefont
  {Y.}~\bibnamefont {Fainman}},\ }\href@noop {} {\enquote {\bibinfo {title}
  {Mesoscopic limit cycles in coupled nanolasers},}\ } (\bibinfo {year}
  {2019}),\ \Eprint {http://arxiv.org/abs/arXiv:1911.10830} {arXiv:1911.10830}
  \BibitemShut {NoStop}%
\bibitem [{\citenamefont {Wang}\ \emph {et~al.}(2015)\citenamefont {Wang},
  \citenamefont {Puccioni},\ and\ \citenamefont {Lippi}}]{Wang2015}%
  \BibitemOpen
  \bibfield  {author} {\bibinfo {author} {\bibfnamefont {T.}~\bibnamefont
  {Wang}}, \bibinfo {author} {\bibfnamefont {G.}~\bibnamefont {Puccioni}}, \
  and\ \bibinfo {author} {\bibfnamefont {G.}~\bibnamefont {Lippi}},\
  }\href@noop {} {\bibfield  {journal} {\bibinfo  {journal} {Sci. Rep.}\
  }\textbf {\bibinfo {volume} {5}},\ \bibinfo {pages} {15858} (\bibinfo {year}
  {2015})}\BibitemShut {NoStop}%
\bibitem [{\citenamefont {Wiersig}\ \emph {et~al.}(2009)\citenamefont
  {Wiersig}, \citenamefont {Gies}, \citenamefont {Jahnke}, \citenamefont
  {A{\ss}mann}, \citenamefont {Berstermann}, \citenamefont {Bayer},
  \citenamefont {Kistner}, \citenamefont {Reitzenstein}, \citenamefont
  {Schneider}, \citenamefont {H{\"o}fling} \emph {et~al.}}]{Wiersig2009}%
  \BibitemOpen
  \bibfield  {author} {\bibinfo {author} {\bibfnamefont {J.}~\bibnamefont
  {Wiersig}}, \bibinfo {author} {\bibfnamefont {C.}~\bibnamefont {Gies}},
  \bibinfo {author} {\bibfnamefont {F.}~\bibnamefont {Jahnke}}, \bibinfo
  {author} {\bibfnamefont {M.}~\bibnamefont {A{\ss}mann}}, \bibinfo {author}
  {\bibfnamefont {T.}~\bibnamefont {Berstermann}}, \bibinfo {author}
  {\bibfnamefont {M.}~\bibnamefont {Bayer}}, \bibinfo {author} {\bibfnamefont
  {C.}~\bibnamefont {Kistner}}, \bibinfo {author} {\bibfnamefont
  {S.}~\bibnamefont {Reitzenstein}}, \bibinfo {author} {\bibfnamefont
  {C.}~\bibnamefont {Schneider}}, \bibinfo {author} {\bibfnamefont
  {S.}~\bibnamefont {H{\"o}fling}},  \emph {et~al.},\ }\href@noop {} {\bibfield
   {journal} {\bibinfo  {journal} {Nature}\ }\textbf {\bibinfo {volume}
  {460}},\ \bibinfo {pages} {245} (\bibinfo {year} {2009})}\BibitemShut
  {NoStop}%
\bibitem [{\citenamefont {Takemura}\ \emph {et~al.}(2012)\citenamefont
  {Takemura}, \citenamefont {Omachi},\ and\ \citenamefont
  {Kuwata-Gonokami}}]{Takemura2012}%
  \BibitemOpen
  \bibfield  {author} {\bibinfo {author} {\bibfnamefont {N.}~\bibnamefont
  {Takemura}}, \bibinfo {author} {\bibfnamefont {J.}~\bibnamefont {Omachi}}, \
  and\ \bibinfo {author} {\bibfnamefont {M.}~\bibnamefont {Kuwata-Gonokami}},\
  }\href {\doibase 10.1103/PhysRevA.85.053811} {\bibfield  {journal} {\bibinfo
  {journal} {Phys. Rev. A}\ }\textbf {\bibinfo {volume} {85}},\ \bibinfo
  {pages} {053811} (\bibinfo {year} {2012})}\BibitemShut {NoStop}%
\bibitem [{\citenamefont {Drummond}\ and\ \citenamefont
  {Walls}(1980)}]{Drummond1980}%
  \BibitemOpen
  \bibfield  {author} {\bibinfo {author} {\bibfnamefont {P.~D.}\ \bibnamefont
  {Drummond}}\ and\ \bibinfo {author} {\bibfnamefont {D.~F.}\ \bibnamefont
  {Walls}},\ }\href {\doibase 10.1088/0305-4470/13/2/034} {\bibfield  {journal}
  {\bibinfo  {journal} {Journal of Physics A: Mathematical and General}\
  }\textbf {\bibinfo {volume} {13}},\ \bibinfo {pages} {725} (\bibinfo {year}
  {1980})}\BibitemShut {NoStop}%
\bibitem [{\citenamefont {Abbaspour}(2015)}]{Abbaspour2015}%
  \BibitemOpen
  \bibfield  {author} {\bibinfo {author} {\bibfnamefont {H.}~\bibnamefont
  {Abbaspour}},\ }\emph {\bibinfo {title} {Noise-Induced Phenomena in
  Collective Spinor Polariton Excitations}},\ \href@noop {} {Ph.D. thesis},\
  \bibinfo  {school} {EPFL}, \bibinfo {address} {The address of the publisher}
  (\bibinfo {year} {2015}),\ \bibinfo {note} {an optional note}\BibitemShut
  {NoStop}%
\bibitem [{\citenamefont {Casteels}\ \emph {et~al.}(2017)\citenamefont
  {Casteels}, \citenamefont {Fazio},\ and\ \citenamefont
  {Ciuti}}]{Casteels2017}%
  \BibitemOpen
  \bibfield  {author} {\bibinfo {author} {\bibfnamefont {W.}~\bibnamefont
  {Casteels}}, \bibinfo {author} {\bibfnamefont {R.}~\bibnamefont {Fazio}}, \
  and\ \bibinfo {author} {\bibfnamefont {C.}~\bibnamefont {Ciuti}},\ }\href
  {\doibase 10.1103/PhysRevA.95.012128} {\bibfield  {journal} {\bibinfo
  {journal} {Phys. Rev. A}\ }\textbf {\bibinfo {volume} {95}},\ \bibinfo
  {pages} {012128} (\bibinfo {year} {2017})}\BibitemShut {NoStop}%
\bibitem [{\citenamefont {Fink}\ \emph {et~al.}(2017)\citenamefont {Fink},
  \citenamefont {Schade}, \citenamefont {Höfling}, \citenamefont {Schneider},\
  and\ \citenamefont {Imamoglu}}]{Fink2017}%
  \BibitemOpen
  \bibfield  {author} {\bibinfo {author} {\bibfnamefont {T.}~\bibnamefont
  {Fink}}, \bibinfo {author} {\bibfnamefont {A.}~\bibnamefont {Schade}},
  \bibinfo {author} {\bibfnamefont {S.}~\bibnamefont {Höfling}}, \bibinfo
  {author} {\bibfnamefont {C.}~\bibnamefont {Schneider}}, \ and\ \bibinfo
  {author} {\bibfnamefont {A.}~\bibnamefont {Imamoglu}},\ }\href {\doibase
  10.1038/s41567-017-0020-9} {\bibfield  {journal} {\bibinfo  {journal} {Nat.
  Phys.}\ } (\bibinfo {year} {2017}),\ 10.1038/s41567-017-0020-9}\BibitemShut
  {NoStop}%
\end{thebibliography}
%

\end{document}